\documentclass[article,aps,prb,english,singlecolumn,notitlepage]{revtex4-1} 
\usepackage[T1]{fontenc} \usepackage[latin1]{inputenc}
 \usepackage{babel} \usepackage{txfonts}
 \usepackage{epsfig,epsf,psfrag} 
\usepackage{graphicx} 
\usepackage{subfigure}
\usepackage{pslatex}
\usepackage{epic,eepic} 
\usepackage{color,pstcol}
\usepackage{pstricks} 
\usepackage{fancyhdr}  \parindent0em  

 \vspace{-10.0cm} 
\begin{document} \title{Are quantum spin Hall edge modes more resilient to disorder, sample geometry and inelastic scattering than quantum Hall edge modes?} 
  \author{Arjun Mani}
 \author{Colin Benjamin} \email{colin.nano@gmail.com}\affiliation{ National institute of Science education \& Research, Bhubaneswar 751005, India }
\begin{abstract}
  On the surface of  2D Topological insulators occur 1D quantum spin Hall edge modes with Dirac like dispersion. Unlike quantum Hall edge modes which occur at high magnetic fields in 2DEGs, the occurrence of quantum spin Hall edge modes is because of spin-orbit scattering in the bulk of the material. These quantum spin Hall edge modes are spin dependent and chiral- opposite spins move in opposing directions. Electronic spin has larger decoherence and relaxation time than charge- in view of this its expected that quantum spin Hall edge modes will be more robust to disorder and inelastic scattering than quantum Hall edge modes which are charge dependent and  spin unpolarized.   However, we notice no such advantage accrues to quantum spin Hall edge modes when subjected to same degree of contact disorder and/or inelastic scattering in similar setups as quantum Hall edge modes. In fact we observe that quantum spin Hall edge modes are more susceptible to inelastic scattering and contact disorder than quantum Hall edge modes. Further, while a single disordered contact has no effect on quantum Hall edge modes it leads to a finite charge Hall current in case of quantum spin Hall edge modes and thus vanishing of pure quantum spin Hall effect. For more than a single disordered contact while quantum Hall states continue to remain immune to disorder, quantum spin Hall edge modes become more susceptible- the Hall resistance for quantum spin Hall effect  changes sign with increasing disorder. In case of many disordered contacts with inelastic scattering included while quantization of Hall edge modes holds, for quantum spin Hall edge modes- a finite charge Hall current still flows. For quantum spin Hall edge modes in the  inelastic scattering regime we distinguish between two cases: with spin-flip and without spin-flip scattering.   Finally, while asymmetry in sample geometry can have a  deleterious effect on quantum spin Hall case it has no impact in quantum Hall case.
 \end{abstract}

\maketitle
\section{Introduction}

Quantum Hall (QH) and quantum spin Hall (QSH) effects have one thing in common- the occurrence of 1D edge modes\cite{goerbig}. Of course there are differences on how they arise- in QH case at high magnetic fields but in QSH case the edge modes arise at zero magnetic fields because of bulk spin orbit effects in 2D topological insulators\cite{sczhang,asboth,hasan}. QH edge modes are chiral with respect to the sample edge- an electron in one of the lower edge modes of  a 2DEG sample will move exactly in the opposite direction to an electron in one of the upper edge modes. QSH edge modes are chiral not only with respect to edge but also spin. For example at upper edge there will be up and down spin edge modes which move in exactly opposite direction. Further these edge modes move in opposite direction to their same spin counterparts at the lower edge.

It is widely known that transport along edge modes in a QH setting is resilient to disorder. In view of this we test whether in a QSH bar the QSH edge modes will show the same resilience or be more resilient to the twin effects of disorder and inelastic scattering- the bane of any phenomena which relies on complete quantum coherence. The expectation is that since QSH edge modes are spin dependent and spin has longer relaxation times than charge, the spin Hall edge modes would be far more robust to disorder and inelastic scattering. However, contrary to expectations we see not only that there is no added advantage of QSH edge modes as against QH edge modes there is rather a disadvantage.  We find QSH edge modes are quite susceptible to disorder and inelastic scattering while QH edge modes aren't. This is of possible to relevance to their use in spintronics and quantum computation applications as also to setups wherein QSH edge modes are utilized to generate Majorana fermions.

The aim of this work is to compare the quantization of Hall and longitudinal resistance seen in an ideal contacted Hall or spin Hall sample and investigate how this quantization is affected by disordered contacts, inelastic scattering and sample geometry. A disordered contact in contradistinction to an ideal contact does not have a transmission probability of one. Further, as sample size increases edge modes will be affected by inelastic scattering, in case inelastic scattering length $l_{in} < L$ (Length of sample). However, edge modes at upper side will not scatter to lower side, what inelastic scattering does is to equilibrate the populations of edge states with each other on same side of the sample. This is the case for inelastic scattering in QH samples. The situation changes in case of QSH samples. Here we have spin-up and spin-down edge modes and equilibration might happen at the same edge between spin up and spin down edge modes in effect via spin flip scattering. In absence of spin flip scattering also edge modes will equilibrate due to inelastic processes like electron-electron scattering or electron-phonon scattering, however this time spin-up edge modes will equilibrate only with spin-up and not spin-down, similarly for spin-down edge modes.   These edge states once equilibrated remain in equilibrium\cite{buti}.  In contrast to an earlier work\cite{jain} which predicted quantized values of conductance in the presence of strong disorder for topological insulator edge modes we show that quantization of longitudinal conductance and Hall conductance is lost even when a single contact is disordered. Of course, one has to caveat the aforesaid statement since Ref.\onlinecite{jain} considers disorder in the sample itself but in the cases we have considered, in this work, the disorder is confined to the contacts only. The effect of random magnetic fluxes on QSH edge modes has been considered earlier\cite{jian} wherein it was concluded that spin Hall edge modes are localized in their presence. Localization of QSH edge modes has also been predicted for non-magnetic disorder in the sample too in Ref.\onlinecite{dolcini}.

  Inelastic scattering however is not restricted to contacts but is all pervasive and comes into picture when the sample length exceeds the inelastic scattering length. Furthermore inelastic scattering may be accompanied by spin-flip scattering too. In the cases we consider below the length of the sample in ideal and single probe  disordered cases are less than inelastic scattering length. Only when we consider the case of a sample with all probe contacts disordered,  we may have length of sample exceeding inelastic scattering length. In our work the term ``probe'' and ``contact'' mean the same- a metallic reservoir. Finally we generalize the results to N terminals (or, probes or, contacts), this will help us in determining whether sample geometry like difference in number of contacts at upper or lower edge will have any bearing on the edge modes.
There will be four parts of the work each with the same sections:
We calculate Hall resistance $R_{H}$, Longitudinal resistance $R_{L}$ and two terminal resistance $R_{14,14}$ for each section in all the parts-
\\
Part 1: QH edge modes: 6 terminal-
  a. ideal case,
  b. single disordered contact, c. two or more disordered contacts,
  d. all disordered contacts with inelastic scattering.
\\
Part 2: QSH edge mode: 6 terminal-
 a. ideal case,
  b. single disordered contact, c. two or more disordered contacts,
  d. all disordered contacts with inelastic scattering (wit spin-flip), e. all disordered contacts with inelastic scattering (without spin-flip) .
 \\
Part 3: Generalize to N terminals, the QH case-
 a. ideal case,
 b. all disordered contacts with  inelastic scattering.
  \\
Part 4: Generalization to N terminals, the QSH case-
 a. ideal case,
  b. all disordered contacts with inelastic scattering.

\section{Quantum  Hall edge modes}
The Landauer-Buttiker formula relating currents and voltages in a multiprobe device is\cite{buti}:
\begin{equation}
I_{i}=\sum_{j} ( G_{ji} V_{i} - G_{ij} V_{j})=\frac{e^{2}}{h} \sum_{j} ( T_{ji} V_{i} - T_{ij} V_{j})
\end{equation}
where $V_{i}$ is the voltage at i$^{th}$ terminal and $I_{i}$ is the current flowing from the same terminal. Here $T_{ij}$ is the transmission from j$^{th}$ to  i$^{th}$ terminal and $G_{ij}$ is the associated conductance. 
\begin{figure*}
  \centering
 \subfigure[Ideal case: Contacts are reflectionless]{ \includegraphics[width=0.3\textwidth]{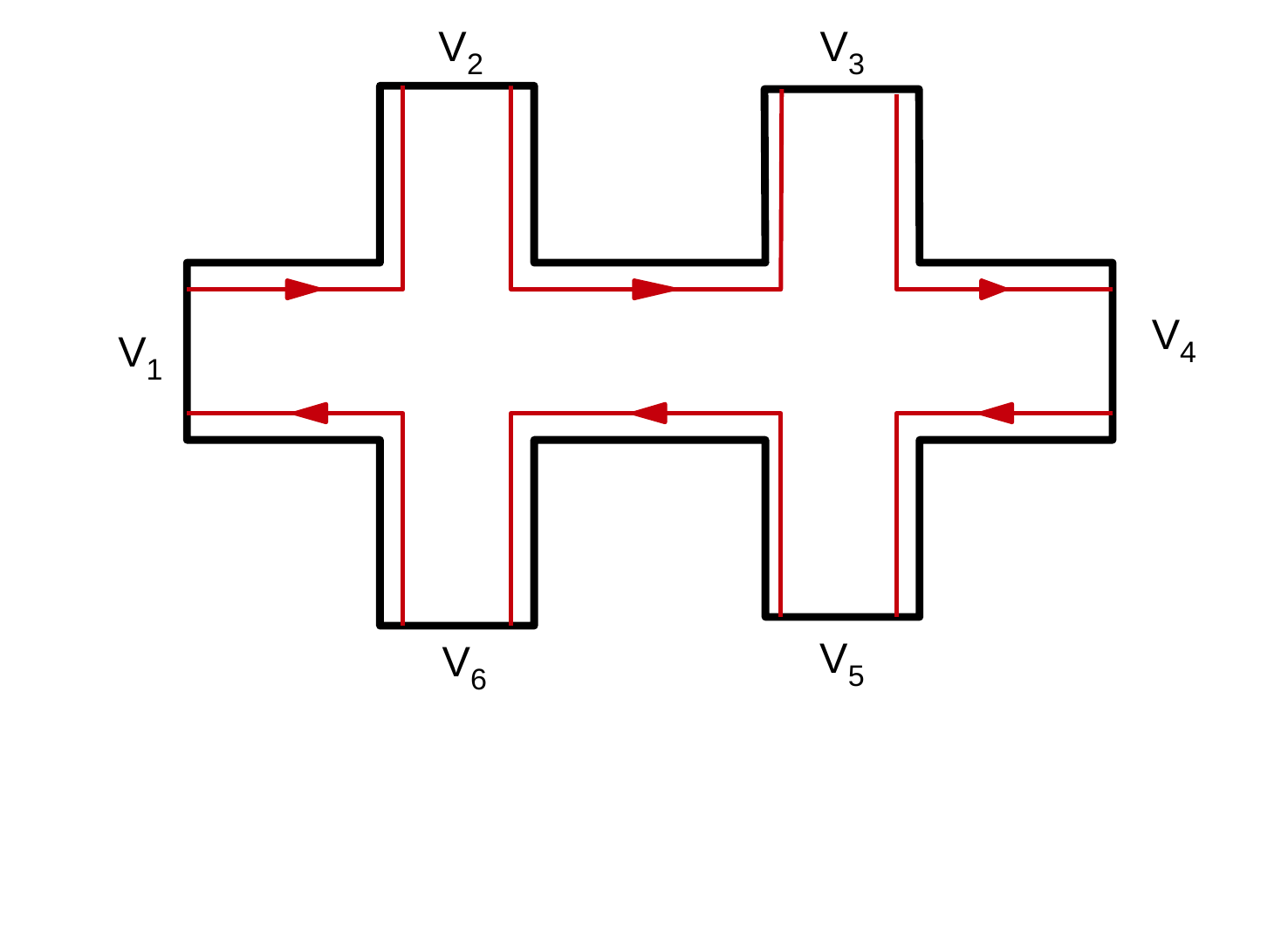}}
  \centering\subfigure  [Single disordered probe: $R_{1}, T_{1}$ represent the reflection and transmission probability of edge modes from and into contact $2$, to represent the effect of disorder in contact 2, an extra edge mode is shown being reflected from contact 2. ]
{    \includegraphics[width=.33\textwidth]{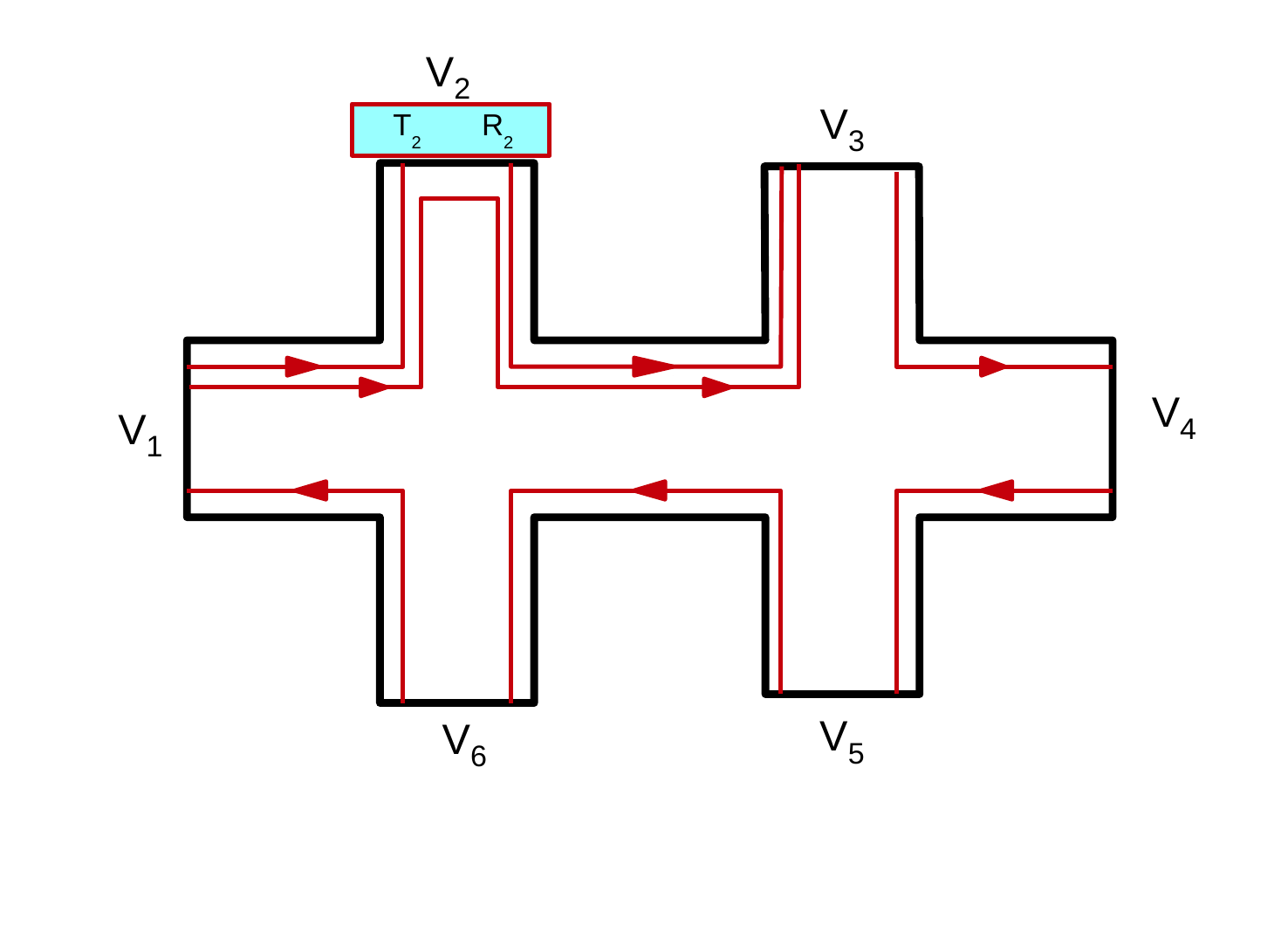}}
 \centering\subfigure [All disordered contacts with inelastic scattering:  Starry blobs indicate equilibration of contact potentials at those places]{ \includegraphics[width=.34\textwidth]{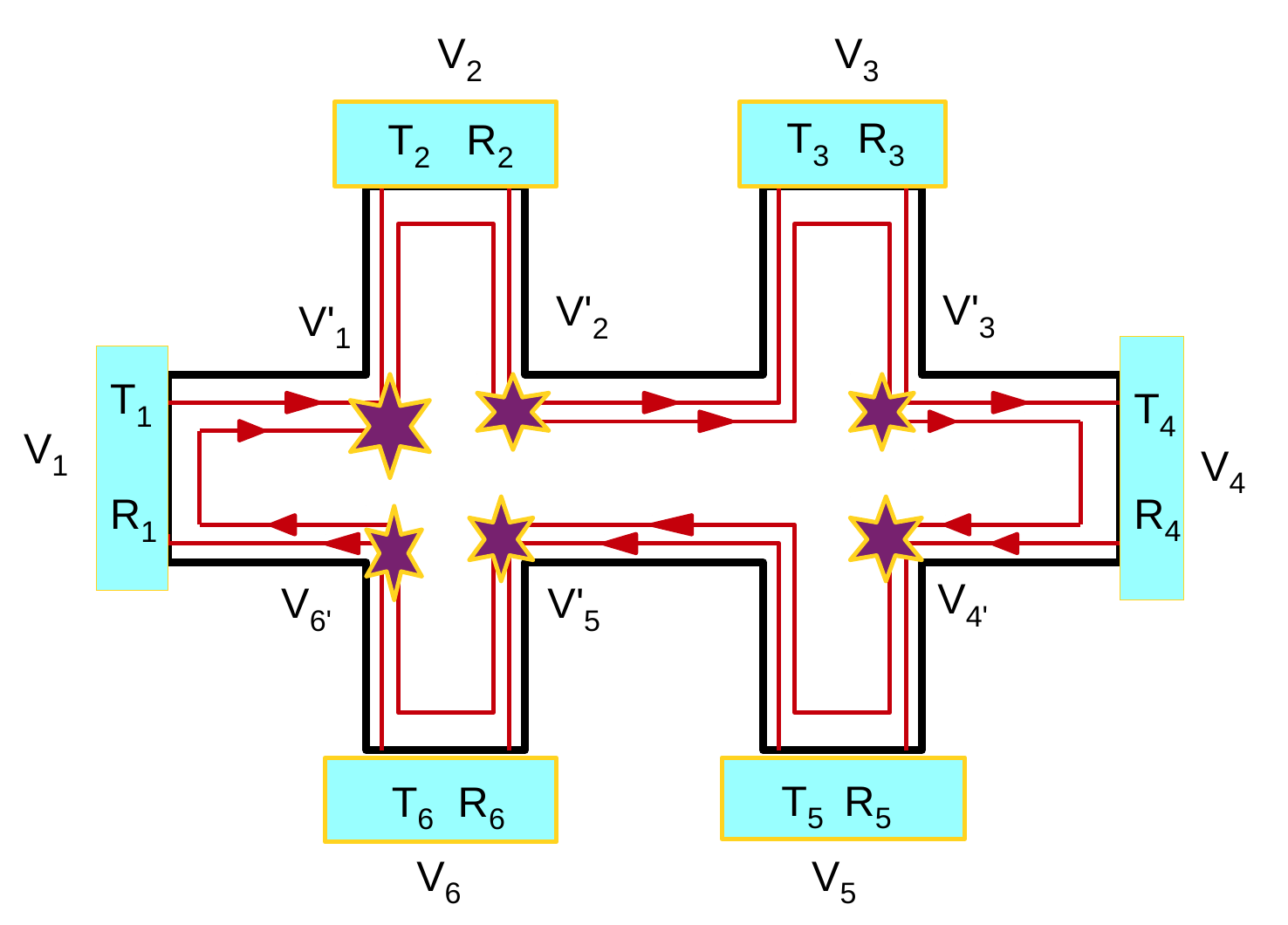}}\caption{Six terminal QH bar showing QH edge modes, there are equal no. of edge modes on both sides of the sample.}
\end{figure*}

\subsection{ QH edge modes- ideal case:}
The ideal case is represented in Fig. 1(a). The current voltage relations can be derived from the conductance matrix below:
\begin{equation}
G_{ij} =-\frac{e^{2} M}{h} \left( \begin{array}{cccccc}
    -1  & 0 & 0 & 0 &0 & 1\\
    1  &-1 & 0 & 0 &0 & 0\\ 0  & 1 & -1 & 0 &0 & 0\\ 0  & 0 & 1 & -1 &0 & 0\\ 0 & 0 & 0 & 1 &-1 & 0\\ 0  & 0 & 0 & 0 &1 & -1\\\end{array} \right)
\end{equation}
M represents the total no. of modes. In setup as shown in Fig. 1, M=1 to avoid clutter. 
Substituting $I_{2}, I_{3}, I_{5},$ and $ I_{6}=0$ and choosing reference potential $V_{4}=0$, we derive $V_{1}=V_{2}=V_{3}$ and $V_{4}=V_{5}=V_{6}=0$. So, the Hall resistance $R_{H}=R_{14,26}=\frac{h}{e^{2}}\frac{1}{M}$, then  longitudinal resistance $R_{L}=R_{14,23}=0$, two terminal resistance  $R_{2T}=R_{14,14}=\frac{h}{e^{2}}\frac{1}{M}$. The Hall conductance is quantized in units of M and thus the longitudinal conductance vanishes and the two terminal conductance too is quantized in units of M.

\subsection{ QH edge modes- single disordered probe}
The disordered probe case is represented in Fig. 1(b). We consider only a single probe to be disordered.
The disorder strengths $D$'s can be written in terms of the T's- the no. of  transmitted edge modes and M- the total no. of edge modes. So, $T_{i}=(1-D_{i})M$ and $R_{i}=D_{i}M$. Depending on the strength of disorder only a fraction of edge modes will be transmitted by the disordered contact rest would be reflected, see Fig 1(b).

The current voltage relations can be derived from the conductance matrix below:
\begin{equation}
G_{ij} =-\frac{e^{2} }{h} \left( \begin{array}{cccccc}
    -M  & 0 & 0 &0 &0 & M\\
    T_{2} &-T_{2} & 0 & 0 &0 & 0\\ R_{2}  & T_{2} & -M & 0 &0 & 0\\ 0  & 0 & M & -M &0 & 0\\ 0 & 0 & 0 & M &-M & 0\\ 0  & 0 & 0 & 0&M& -M\\\end{array} \right) 
\end{equation}
Here, $M$ is total no. of edge modes, $T$'s are the no. of transmitted edge modes into a contact and $R$'s are no. of reflected edge modes.
Substituting $I_{2}, I_{3}, I_{5},$ and $ I_{6}=0$ and choosing reference potential $V_{4}=0$, we derive $V_{1}=V_{2}=V_{3}$, and $V_{4}=V_{5}=V_{6}=0$. So, the Hall resistance $R_{H}=R_{14,26}=\frac{h}{e^{2}}\frac{1}{M}$,   longitudinal resistance $R_{L}=R_{14,23}=0$, two terminal resistance  $R_{2T}=R_{14,14}=\frac{h}{e^{2}}\frac{1}{M}.$
 We see the resistance characteristics are completely independent of the strength of disorder. Similar to the case of ideal contacts,  the Hall conductance is quantized in units of M and thus the longitudinal conductance vanishes and the two terminal conductance too is quantized in units of M.

\subsection {QH edge modes- two or more disordered probes} 
The  case of more than a single disordered probes is  extension of that represented in Fig. 1(b) of a single disordered probe with all probes  having finite disorder leading to non-zero  reflection probability for  edge modes from those probes.  Herein, we consider all the contacts to be disordered in general. The current voltage relations can be derived from the conductance matrix below:
\begin{equation}
G_{ij} =-\frac{e^{2} M}{h} \left( \begin{array}{cccccc}
    -T^{11}  & T^{12} & T^{13} & T^{14} & T^{15} & T^{16}\\
    T^{21}  & -T^{22} & T^{23} & T^{24} & T^{25} & T^{26}\\ 
    T^{31}  & T^{32} & -T^{33} & T^{34} & T^{35} & T^{36}\\
     T^{41}  & T^{42} & T^{43} & -T^{44} & T^{45} & T^{46}\\ 
     T^{51}  & T^{52} & T^{53} & T^{54} & -T^{55} & T^{56}\\ 
     T^{61}  & T^{62} & T^{63} & T^{64} & T^{65} & -T^{66}\\ \end{array} \right)
\end{equation}
M represents the total no. of modes. In setup as shown in Fig. 1b, M=1 to avoid clutter and only one mode is shown. 
In the above matrix $T^{15}$ say is defined as the total transmission probability from contact $5$ to 
$1$ and can be written explicitly as 
\begin{eqnarray}
T^{15}&=&{(1-D_{5})D_{6}(1-D_{1})M+(1-D_{5})D_{6}^2D_1D_2D_3D_4D_5(1-D_{1})M+...} \nonumber\\
{\mbox or}, T^{15}&=&{(1-D_{5})D_{6}(1-D_{1})M[1+D_1D_2D_3D_4D_5D_6+...]}   ={\frac{(1-D_{5})D_{6}(1-D_{1})M}{1-D_1D_2D_3D_4D_5D_6} }
\end{eqnarray}
this can be understood as follows- $D_{i}$ being the strength of disorder in contact $i$. An electron starting from contact 5 has probability $1-D_{5}$ to be transmitted since we are interested in the probability of its  reaching contact 1, it has to be reflected from contact 6 with probability $D_6$ and finally it is transmitted to contact 1 with probability $1-D_1$, However, this is the shortest of the many paths possible for an electron starting from 5 to reach 1, another path can be that of an electron starting from contact 5, has probability $1-D_{5}$ to be transmitted, since we are interested in the probability of its  reaching contact 1, it has to be reflected from contact 6 with probability $D_6$ and then reflected from contact 1 with probability  $D_1$, similarly with probability $D_{2}$ it will be reflected from contact 2, with probability $D_3$ from contact 3, with probability $D_4$ from contact 4, with probability $D_5$ from contact 5 again get reflected with probability $D_6$ from contact 6 and finally get transmitted into contact 1, this is the second shortest path possible, similarly one can sum over all the other paths leading to an infinite series, which can be summed to yield the total probability per mode for transmission from contact 5 to 1 is as in Eq.~5. Similarly all other transmission probabilities can be derived.   Substituting $I_{2}, I_{3}, I_{5},$ and $ I_{6}=0$ as these are voltage probes and choosing reference potential $V_{4}=0$, we solve the above matrix and calculate  the Hall, longitudinal and 2-terminal resistances. The Hall resistance $R_{H}=R_{14,26}=\frac{h}{e^{2}}\frac{1}{M}$, longitudinal resistance $R_{L}=R_{14,23}=0$, two terminal resistance  $R_{2T}=R_{14,14}=\frac{h}{e^{2}M}\frac{1-D_{1}D_{4}}{(1-D_{1})(1-D_{4})}$ Figs. 3(b-d). The Hall and longitudinal  conductances are all ideally quantized and do not deviate from their ideal results however the 2terminal resistance does deviate when more than one contact  is disordered as in Fig. 3(b-d).

\subsection{ QH edge modes- all disordered  contacts with inelastic scattering:}
The case of QH edge modes in presence of all disordered contacts and with inelastic scattering included has been dealt with before in Ref.\cite{buti,nikolajsen}. We can look at the Figure 1(c) where we consider that the length between disordered contacts is larger than inelastic scattering length. On the occurrence of an inelastic scattering event, the edge states originating from different contacts with different energies are equilibrated to a common potential. In Fig. 1(c), one can see that electrons coming from contact 1 and 6 are equilibrated to potential $V_{1}^{\prime}$. If as before contacts 1 and 4 are chosen to be the current contacts then no current flows into the other voltage probe contacts. Lets say a current $\frac{e^2}{h} T_{2} V_{1}^{\prime}$ enters contact 2 while current $\frac{e^2}{h} T_{2} V_{2}$ leaves contact 2, and since contact 2 is a voltage probe net current has to be zero, implying $V_{2}=V_{1}^{\prime}$. The same thing happens at contact 3 and along the lower edge where states are equilibrated to $V_{4}^{\prime}$.

Now we write the current voltage relations in continuous fashion, eschewing our earlier method of writing it in matrix form to avoid clutter as there are not only the 6 potentials $V_{1}-V_{6}$, we also have the equilibrated potentials $V_{1}^{\prime}-V_{6}^{\prime}$.

\begin{eqnarray}
I_{1}&=& T_{1}(V_{1}-V_{6}^{\prime}),\nonumber\\
I_{2}&=& T_{2}(V_{2}-V_{1}^{\prime}),\nonumber\\
I_{3}&=& T_{3}(V_{3}-V_{2}^{\prime}),\nonumber\\
I_{4}&=& T_{4}(V_{4}-V_{3}^{\prime}),\nonumber\\
I_{5}&=& T_{5}(V_{5}-V_{4}^{\prime}),\nonumber\\
I_{6}&=& T_{6}(V_{6}-V_{5}^{\prime}).
\end{eqnarray}
By putting the condition of net current into voltage probe contacts $2,3,4,5$ to be zero we get the following relations between the contact potentials: $V_{2}=V_{1}^{\prime}, V_{3}=V_{2}^{\prime}, V_{5}=V_{4}^{\prime}, \mbox{and  } V_{6}=V_{5}^{\prime} $.
Further, due to the equilibration the net current just out of contact 2 is the sum  $\frac{e^2}{h} (T_{2} V_{2}+R_{2} V_{1}^{\prime})$ and this should be equal to   $\frac{e^2}{h} M V_{2}^{\prime}$ which is the equilibrated potential due to inelastic scattering at contact 3.
Thus, $\frac{e^2}{h} (T_{2} V_{2}+R_{2} V_{2})=\frac{e^2}{h} M V_{2}^{\prime}$, or $V_{2}=V_{2}^{\prime}$, as $T_{2}+R_{2}=M$ the total no. of edge modes in the system. Thus all the upper edges are equilibrated to same potential $V_{1}^{\prime}=V_{2}=V_{2}^{\prime}=V_{3}=V_{3}^{\prime}$.  Similarly for the equilibrated potentials at the lower end we get $V_{4}^{\prime}=V_{5}=V_{5}^{\prime}=V_{6}=V_{6}^{\prime}$. Using these relations between primed and unprimed potentials  we derive  the Hall resistance $R_{H}=R_{14,26}=\frac{h}{e^{2}}\frac{1}{M}$,   longitudinal resistance $R_{L}=R_{14,23}=0$ and two terminal resistance  $R_{2T}=R_{14,14}=\frac{h}{e^{2}}\frac{M^{2}-R_{1}R_{4}}{MT_{1}T_{4}}$. The QH edge mode Hall conductance apart from the two terminal case remains quantized even when all contacts are disordered with inelastic scattering included.
\begin{figure*}
  \centering \subfigure[ Ideal case: Contacts are reflection-less]{ \includegraphics[width=0.3\textwidth]{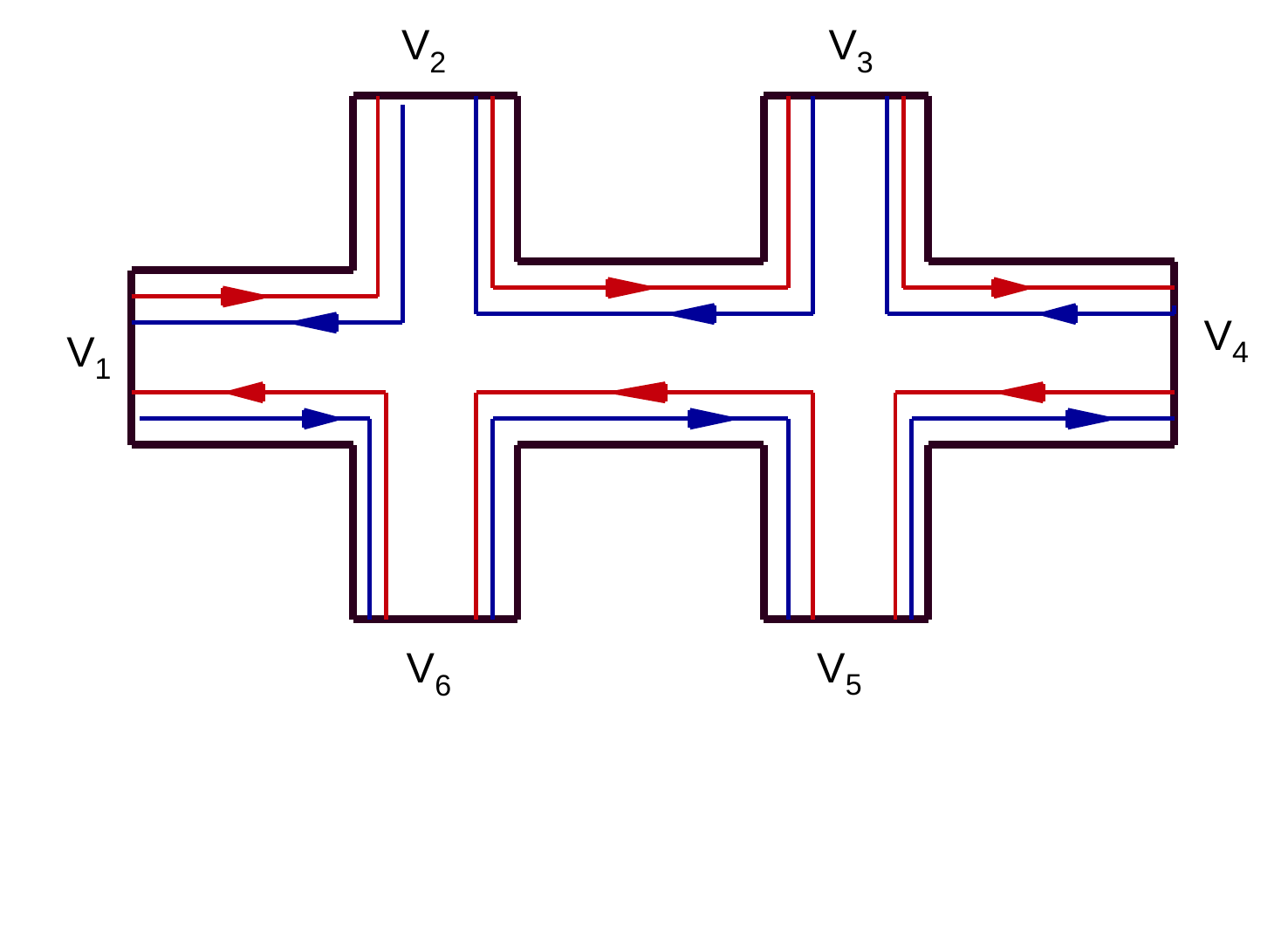}}
 \centering    \subfigure[Single disordered probe: $R_{1}, T_{1}$ represent the reflection and transmission probability of edge modes from and into contact $1$, an extra edge mode is shown to represent the effect of disorder in contact 2]{ \includegraphics[width=.33\textwidth]{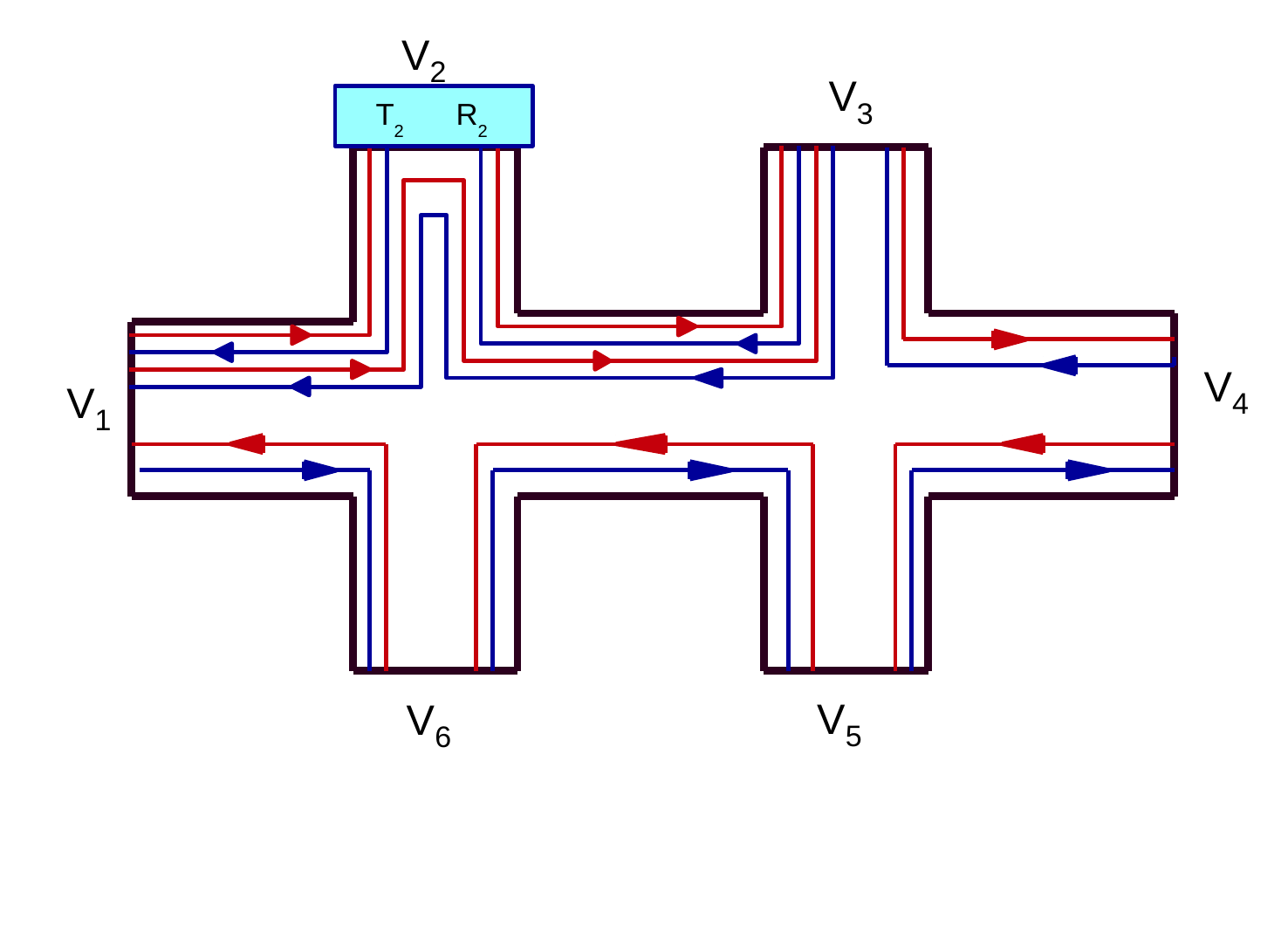}}
 \centering \subfigure[All disordered contacts with inelastic scattering: Starry blobs indicate equilibration of contact potentials at those places]{\includegraphics[width=.33\textwidth]{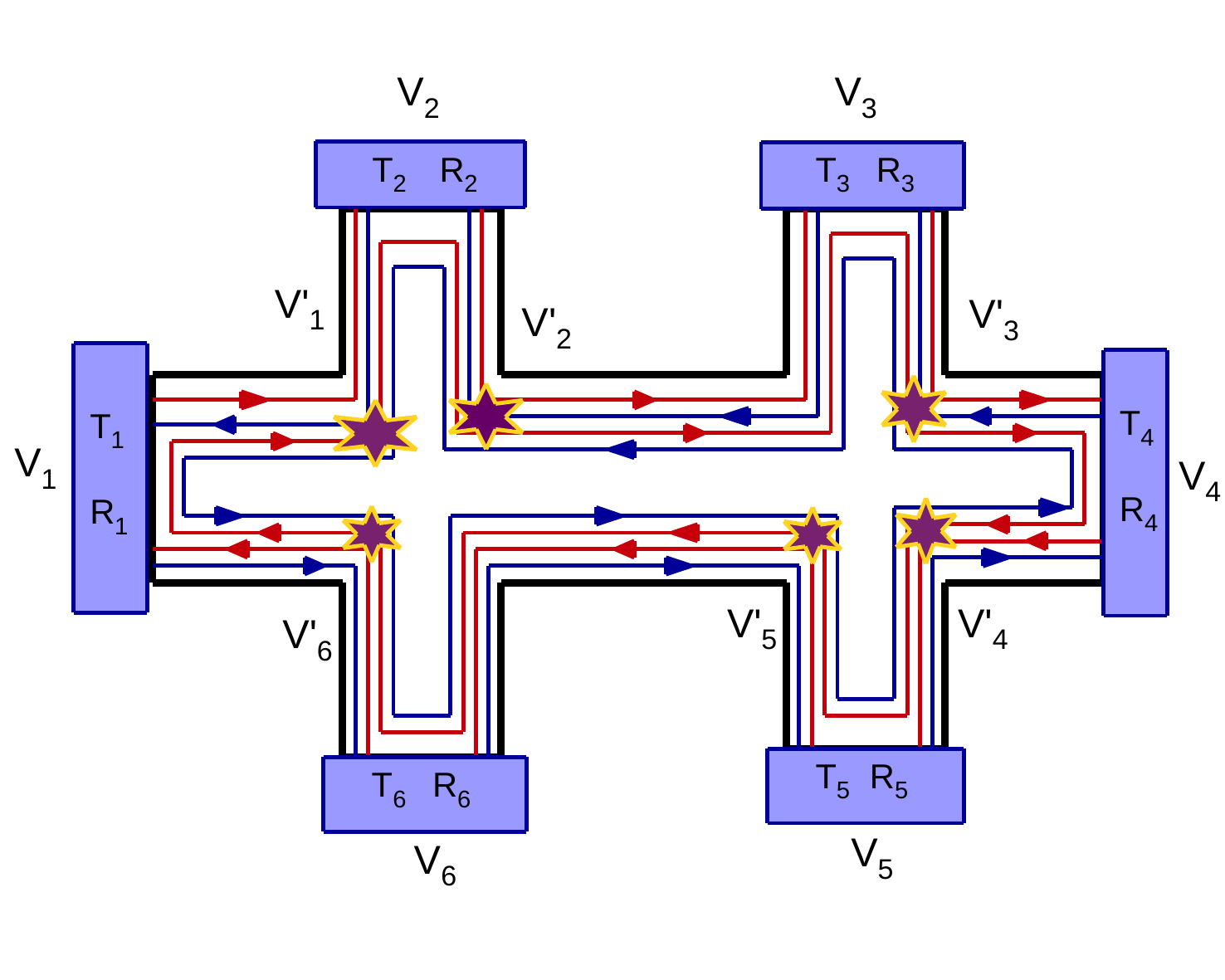}}
 \caption{Six terminal QSH bar showing QSH edge modes. These edge modes differ from their QH counterparts since these are spin polarized and chiral not only with respect to the edges (like QH edge modes) but also with respect to spin: Spin-up and Spin-down edge modes move in opposite directions.}
\end{figure*} 
\section{QSH edge modes}
QH edge modes are chiral and describe electron motion along sample boundaries. Each boundary has an integer number of states or modes, all of which carry charge in the same direction. QSH edge modes one the other hand are spin dependent as also chiral with the additional fact that opposite spins move in opposite directions at one edge\cite{buti-sci}. 
The Landauer-Buttiker formula relating currents and voltages in a multi probe device have been extended to the case of  QSH  edge modes in Ref.\cite{sanvito,chulkov}:
\begin{equation}
I_{i}=\sum_{j} ( G_{ji} V_{i} - G_{ij} V_{j})=\frac{e^{2}}{h} \sum_{j=1}^{N} ( T_{ji} V_{i} - T_{ij} V_{j})
\end{equation}
where $V_{i}$ is the voltage at i$^{th}$ terminal and $I_{i}$ is the current flowing from the same terminal. Here $T_{ij}$ is the transmission from j$^{th}$ to  i$^{th}$ terminal and $G_{ij}$ is the associated conductance. N denotes the no. of terminals/contacts in the system. We first consider N=6 and then generalize it to N terminal case as above.
\subsection{ QSH edge modes- ideal case }
The ideal case is represented in Fig. 2(a). The current voltage relations can be derived from the conductance matrix below:
\begin{equation}
G_{ij} =-\frac{e^{2} M}{h} \left( \begin{array}{cccccc}
    -2 & 1& 0 & 0 &0 & 1\\
    1  &-2 & 1 & 0 &0 & 0\\ 0  & 1 & -2 & 1 &0 & 0\\ 0  & 0 & 1 & -2 &1 & 0\\ 0 & 0 & 0 & 1 &-2 & 1\\ 1  & 0 & 0 & 0 &1 & -2\\\end{array} \right),
\end{equation}

Substituting $I_{2}, I_{3}, I_{5},$ and $ I_{6}=0$ and choosing reference potential $V_{4}=0$, we derive $V_{3}=V_{2}/2=V_{1}/3$ and $V_{5}=V_{6}/2=V_{1}/3$. So, the charge Hall resistance $R_{H}=R_{14,26}=0$ obviously, then  longitudinal resistance is $R_{L}=R_{14,23}=\frac{h}{e^{2}}\frac{1}{2M}$, two terminal resistance  $R_{2T}=R_{14,14}=\frac{h}{e^{2}}\frac{3}{2M}$.  Thus, charge Hall conductance is zero, longitudinal conductance is quantized in units of $2M$ while two terminal conductance is quantized in units of $\frac{2}{3}M$. 

\subsection{ QSH edge modes- single disordered probe}
This case is represented in Fig. 2(b), only a single probe is disordered.
The current voltage relations can be derived from the conductance matrix below:
\begin{equation}
G_{ij} =-\frac{e^{2} }{h} \left( \begin{array}{cccccc}
    -2M & T_{2}& R_{2} & 0 &0 & M\\
    T_{2} &-2T_{2} & T_{2} & 0 &0 & 0\\ R_{2}  & T_{2} & -2M & M &0 & 0\\ 0 & 0 & M & -2M &M & 0\\ 0 & 0 & 0 & M &-2M & M\\ M  & 0 & 0 & 0 &M & -2M\\\end{array} \right),
\end{equation}

Substituting $I_{2}, I_{3}, I_{5},$ and $ I_{6}=0$ and choosing reference potential $V_{4}=0$, we derive $V_{5}=V_{6}/2=V_{1}/3$. So, the Hall resistance $R_{H}=R_{14,26}=\frac{h}{2 e^{2} M}\frac{D_{2}}{3+2D_{2}}$, then  longitudinal resistance $R_{L}=R_{14,23}=\frac{h}{2 e^{2} M}\frac{3}{3+2D_{2}}$, two terminal resistance  $R_{2T}=R_{14,14}=\frac{h}{2 e^{2} M}\frac{9+3D_{2}}{3+2D_{2}}$. Thus all of the calculated conductances the Hall , longitudinal and two terminal lose their quantization and are dependent on disorder. Notice that these quantities are all influenced by disorder in contrast to the QH case in which they are immune to disorder.
\begin{figure*}
 \centering    \subfigure[Resistance  vs. Disorder $D$ for single probe disorder.]{ \includegraphics[width=.45\textwidth]{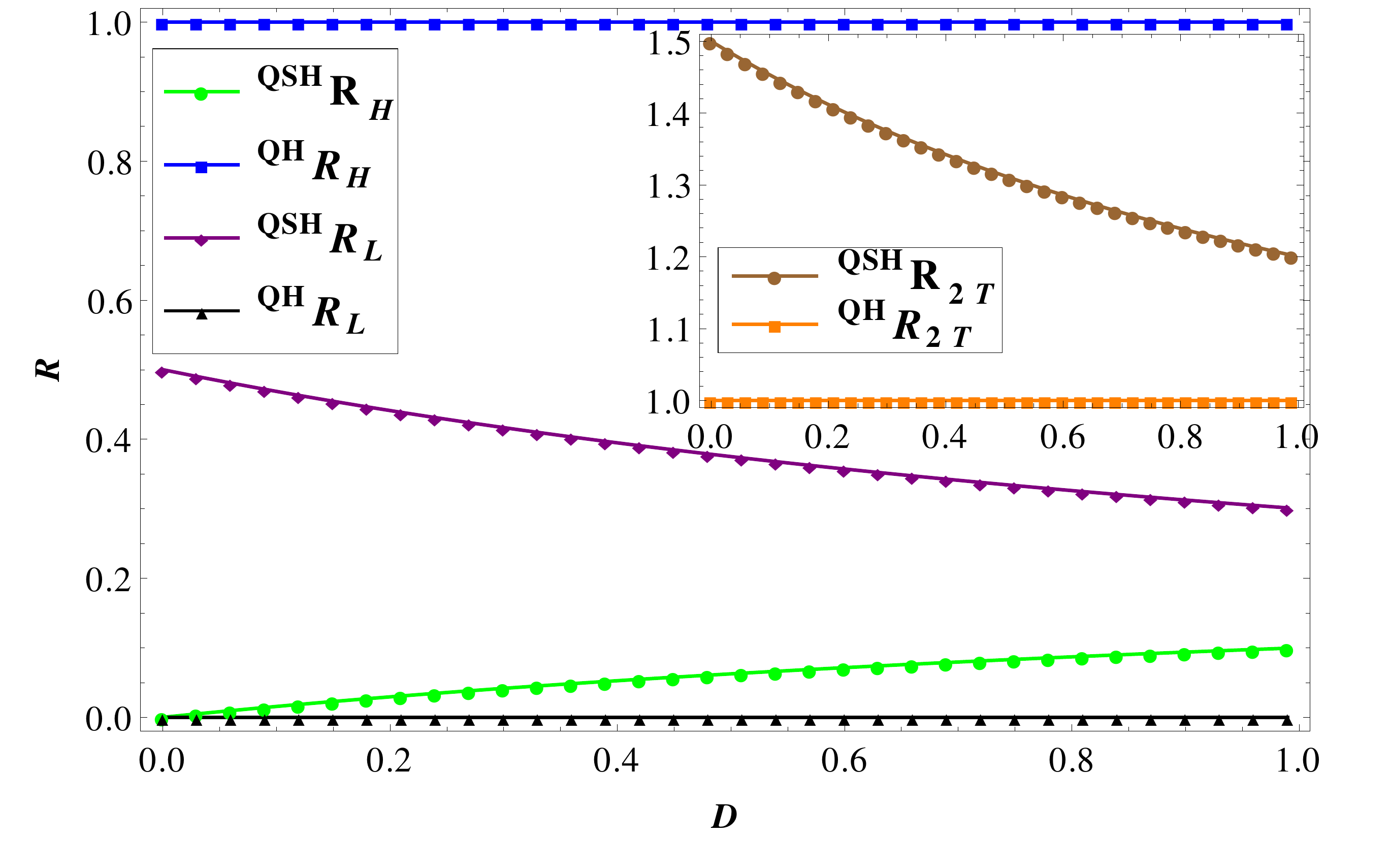}}
 \centering    \subfigure[$R_{H}, R_{L}$ and $R_{2T}$  vs. Disorder $D=D_{2}$ for all disordered probes in case of QH and QSH cases without inelastic scattering $D_{1}=D_{4}=0, D_{3}=0.2,D_{5}=0.5, D_{6}=0.3$ ]{ \includegraphics[width=.45\textwidth]{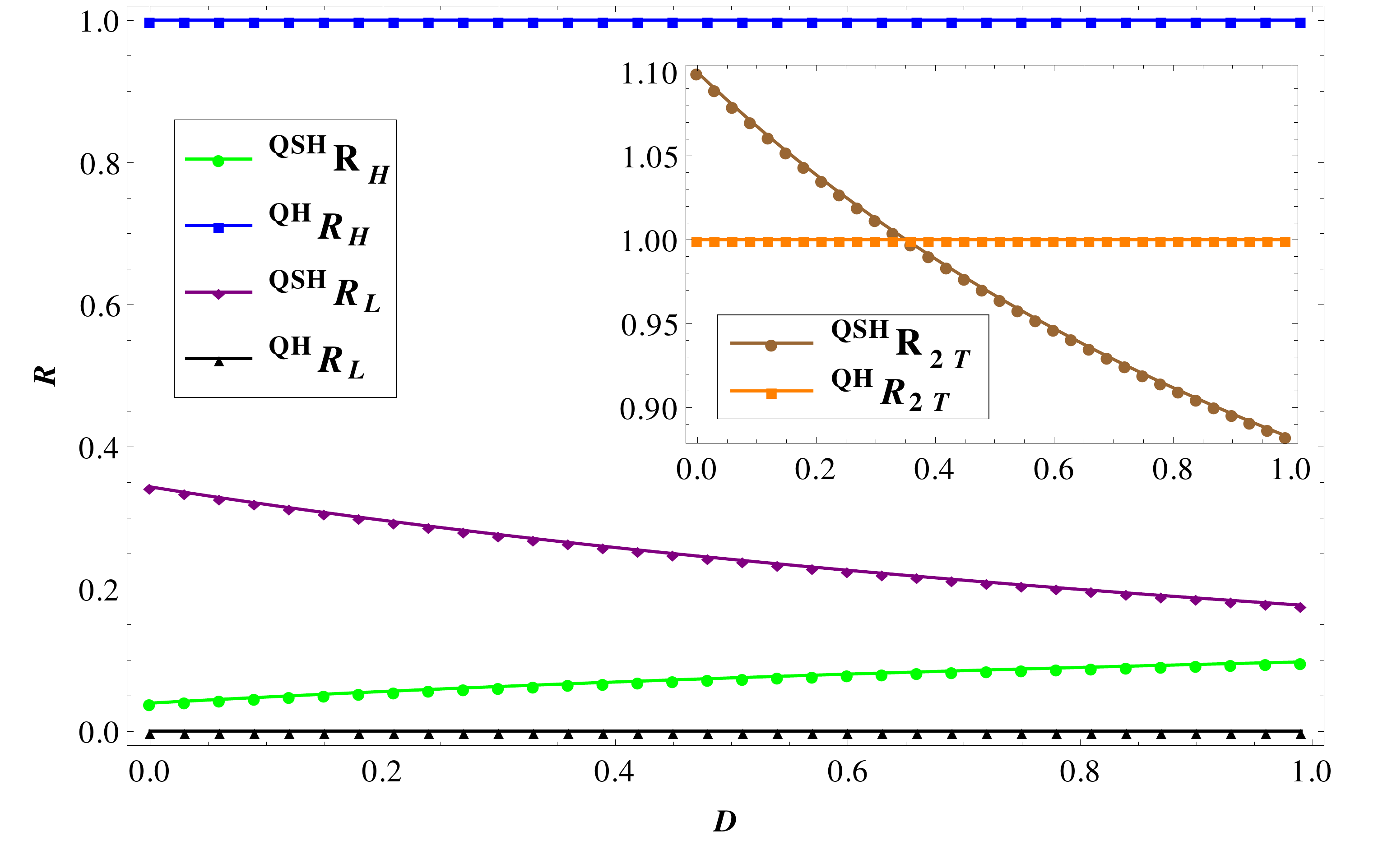}}
 \centering \subfigure[$R_{H}, R_{L}$ and $R_{2T}$  vs. Disorder $D=D_{5}$ for two disordered probes in case of QH and QSH cases without inelastic scattering $D_{1}=D_{6}=D_{2}=D_{3}=0, D_{4}=0.5$ ]{ \includegraphics[width=.45\textwidth]{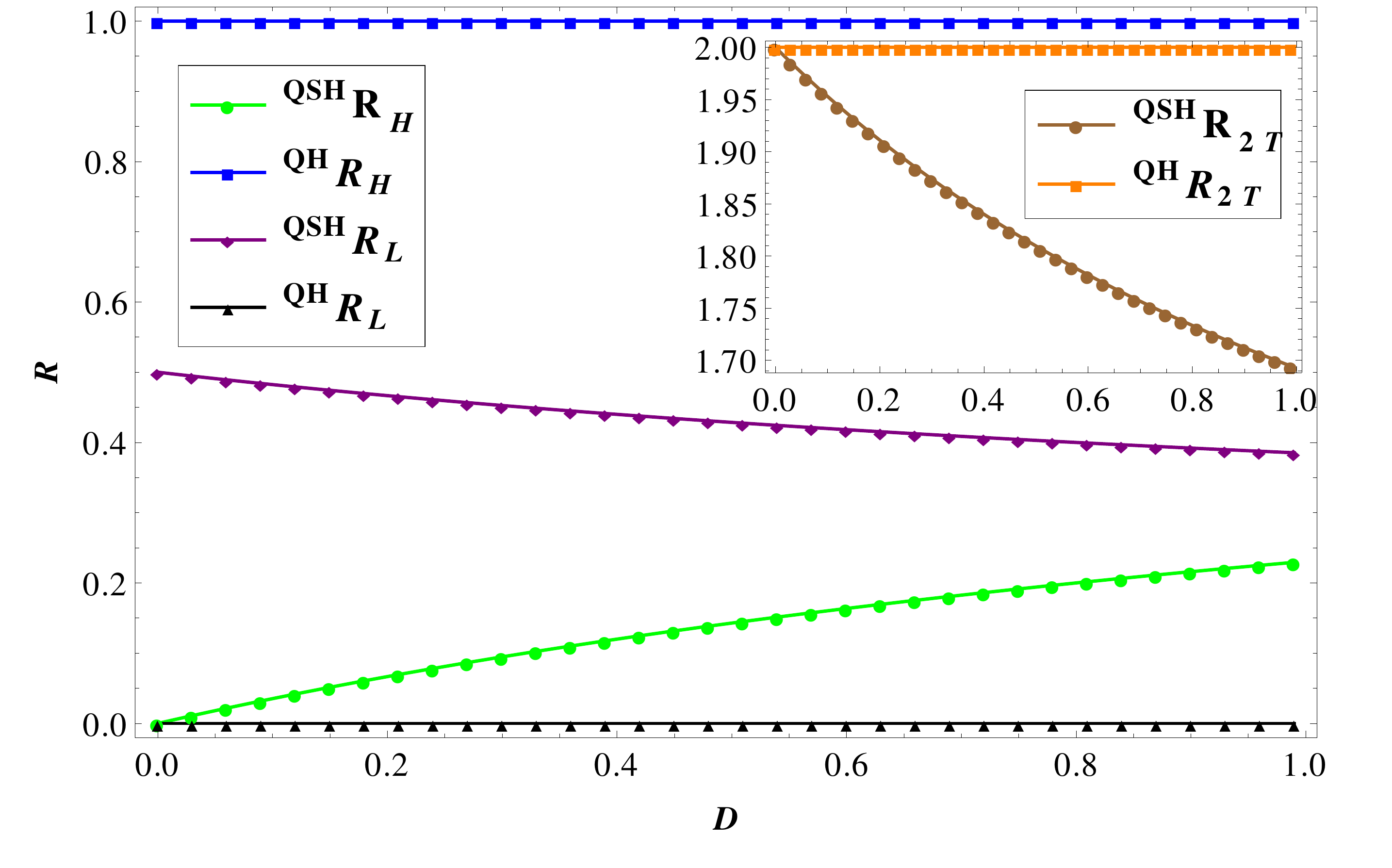}}
 \centering \subfigure[$R_{H}, R_{L}$ and $R_{2T}$  vs. Disorder $D=D_{5}$ for all disordered probes in case of QH and QSH cases without inelastic scattering $D_{1}=D_{6}=0.2, D_{2}=D_{3}=D_{4}=0.5,$ ]{ \includegraphics[width=.45\textwidth]{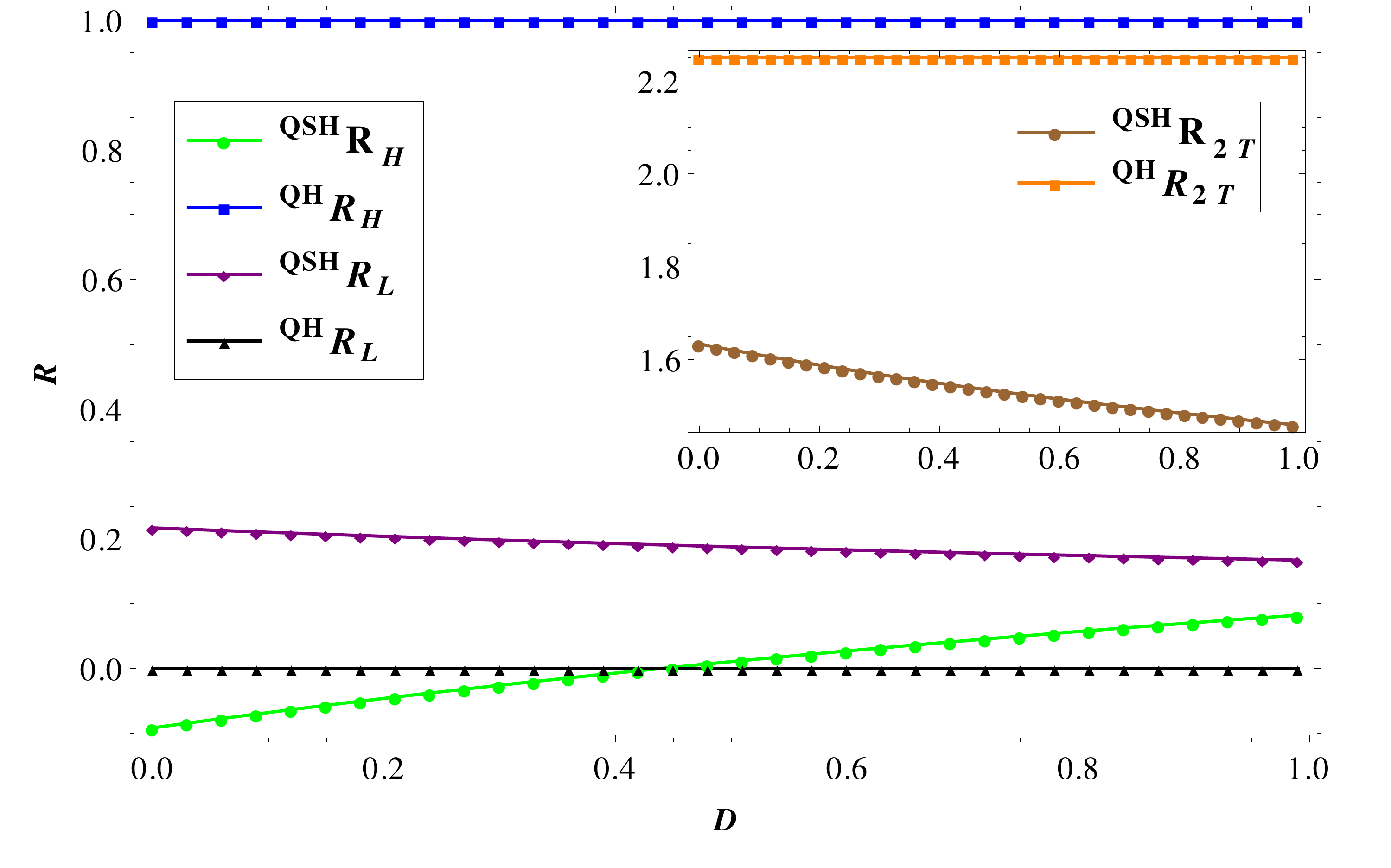}}
 \centering    \subfigure[Hall Resistance  vs. Disorder $D_{1}=D_{4}=D$ for all probe disorder ($D_{2}=D_{3}=0.5$ and $D_{5}=D_{6}=0.8$) with inelastic scattering ]{ \includegraphics[width=.45\textwidth]{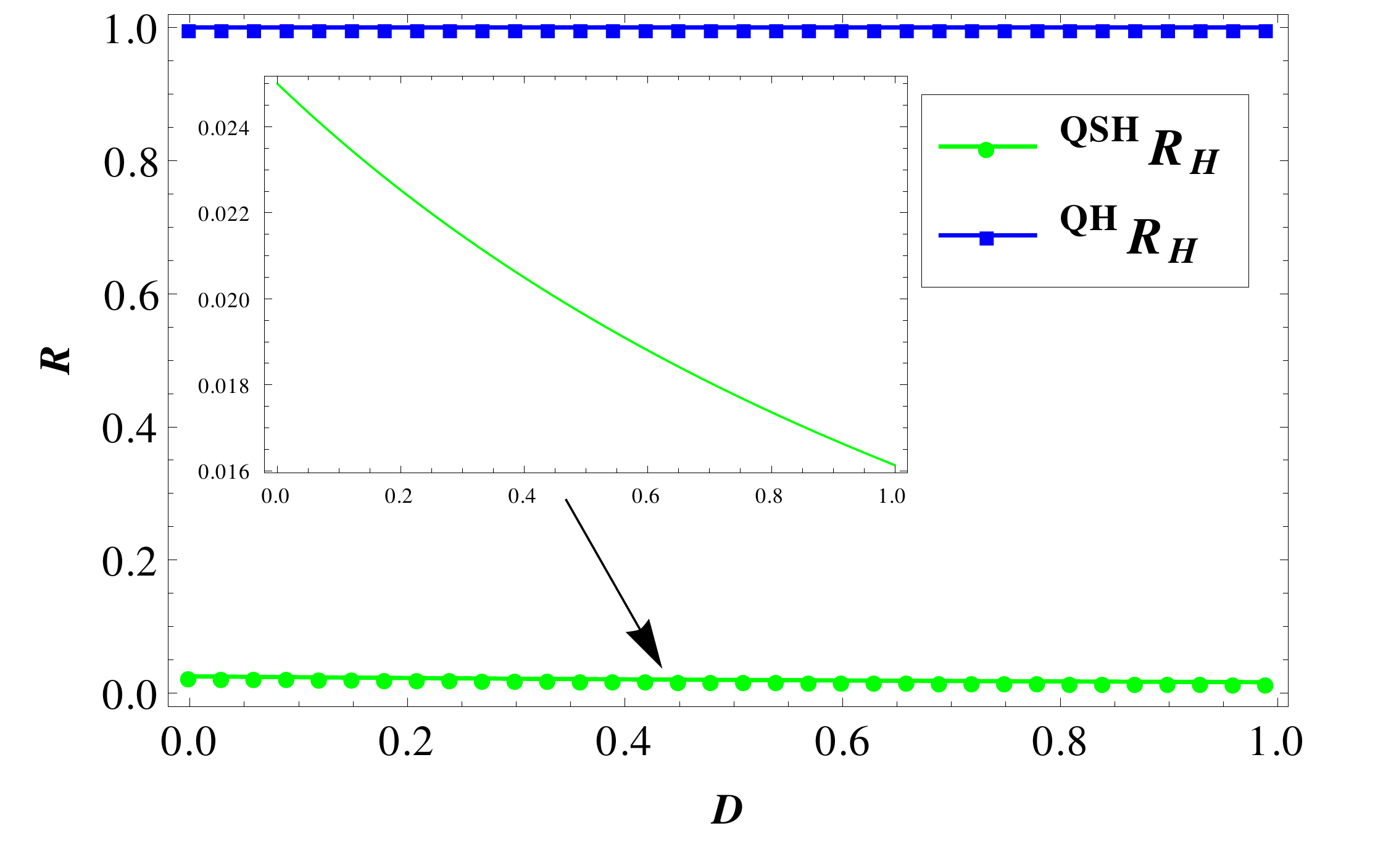}}
\centering    \subfigure[Longitudinal Resistance  vs. Disorder $D_{1}=D_{4}=D$ for all probe disorder ($D_{2}=D_{3}=0.5$ and $D_{5}=D_{6}=0.8$)               
  with inelastic scattering ]{ \includegraphics[width=.45\textwidth]{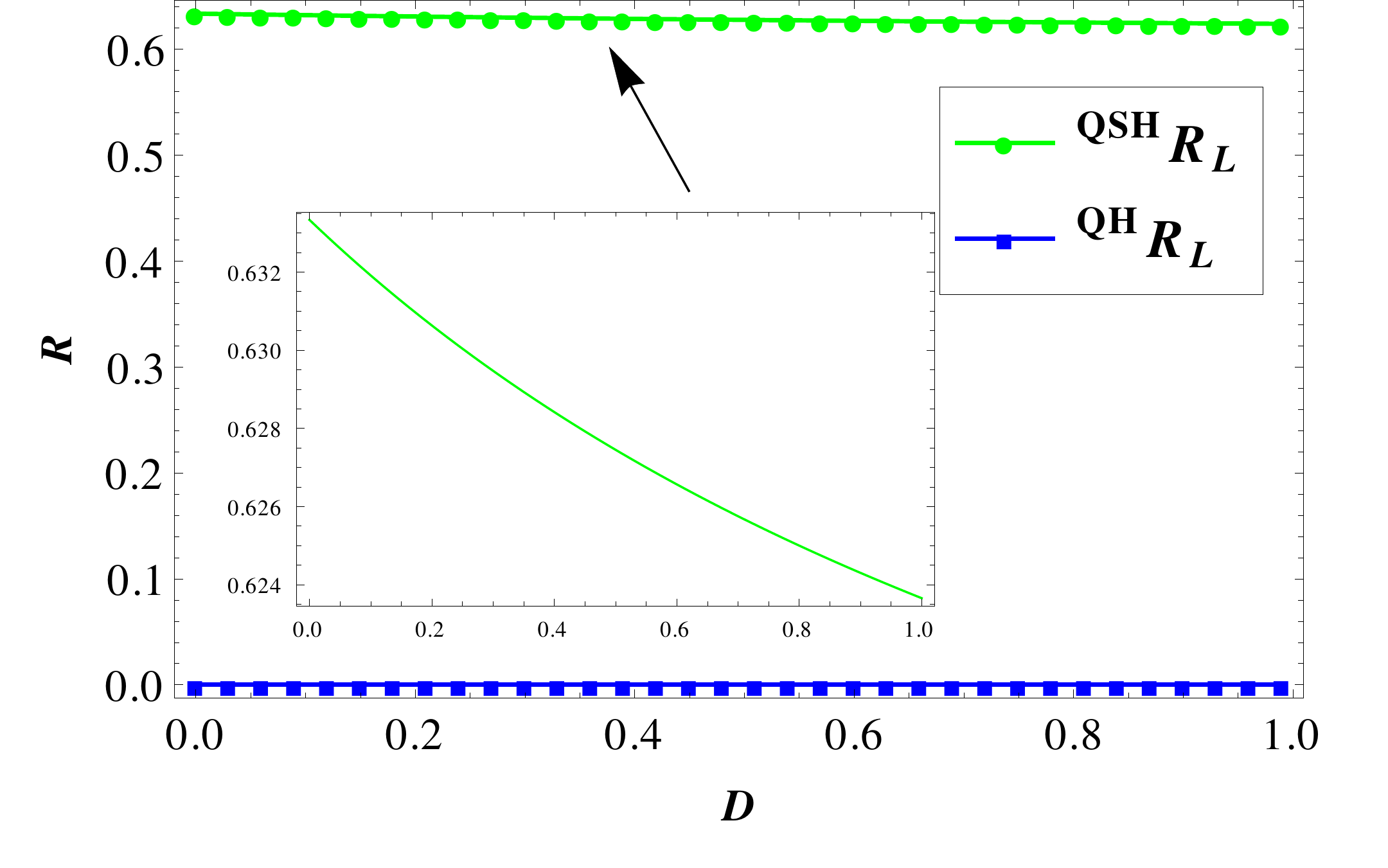}} 
\caption{$R_{H}, R_{L}, R_{2T}$ vs. Disorder in probe (a) $D_{1}$ (b) $D_{2}$ (c) $D_{5}$ (d) $D_{5}$, (e) $D_{1}$ and (f) $D_{1}$. }
\end{figure*} 

\subsection {QSH edge modes- two or more disordered probes} 
The  case for all disordered probes is an extension of the case of single disordered probes as represented in Fig. 2(b). 
Herein we consider all the contacts to be disordered in general. The current voltage relations can be derived from the conductance matrix below:
\begin{equation}
G_{ij} =-\frac{e^{2} M}{h} \left( \begin{array}{cccccc}
    -T^{11}  & T^{12} & T^{13} & T^{14} & T^{15} & T^{16}\\
    T^{21}  & -T^{22} & T^{23} & T^{24} & T^{25} & T^{26}\\ 
    T^{31}  & T^{32} & -T^{33} & T^{34} & T^{35} & T^{36}\\
     T^{41}  & T^{42} & T^{43} & -T^{44} & T^{45} & T^{46}\\ 
     T^{51}  & T^{52} & T^{53} & T^{54} & -T^{55} & T^{56}\\ 
     T^{61}  & T^{62} & T^{63} & T^{64} & T^{65} & -T^{66}\\ \end{array} \right)
\end{equation}
M represents the total no. of modes. In setup as shown in Fig. 1b, M=1 to avoid clutter and only one mode is shown. 
In the above matrix $T^{15}$-the total transmission probability from contact $5$ to 
$1$, in contradistinction to that for QH case  and can be written explicitly as
 \begin{equation}
T^{15}=\frac{[(1-D_{5})D_{6}(1-D_{1})+(1-D_{5})D_{4}D_{3}D_{2}(1-D_{1})]M}{1-D_1D_2D_3D_4D_5D_6}
\end{equation}
the reason being there are two spin polarized edge modes which are moving in opposite directions, the up spin polarized edge mode contributes to $T^{15}$ via the first term while the down spin polarized edge mode contributes via the second term. So the total probability per mode for transmission from contact 5 to 1 is as defined above. Similarly all other transmission probabilities occurring in the above matrix can be explained.  Substituting $I_{2}, I_{3}, I_{5},$ and $ I_{6}=0$ as these are voltage probes and choosing reference potential $V_{4}=0$, we solve the above matrix and calculate the Hall, longitudinal and 2Terminal resistances. Since the expressions for these are quite large we only analyze them via plots as in Fig. 3(b-d). As previously noted for QH case in section II C, we see that the difference between QH and spin Hall is also quite stark when it comes to more than one disordered contact. In QH case, while the Hall and longitudinal resistances do not deviate from ideal quantized values for QSH case these deviate from their ideal quantized values.  In fact for a particular choice as in Fig. 3 (d), the Hall current for spin Hall edge modes not only is finite but it changes sign indicating the complete breakdown of the spin Hall effect via disorder. Further for other choices of disorder probes as in Fig 3 (b) and 3(c) we see while Hall, longitudinal and 2-terminal resistance for QH case is quantized while the same quantities for QSH case deviate from their ideal quantized values indicating that the QSH  edge modes are much more fragile than QH edge modes.
\subsection{QSH edge modes- all disordered probes with inelastic scattering ( with spin flip)}
The case of QSH edge modes in presence of completely disordered contacts and with inelastic scattering included can be understood by extending the approach towards QH edge modes. We can look at the Figure 2(c) where we consider that the length between disordered contacts is larger than inelastic scattering length. On the occasion of an inelastic scattering event happening the edge states originating from different contacts with different energies are equilibrated to a common potential as in QH case. In Fig. 2(c), one can see that electrons coming from contact 1 and 6 are equilibrated to potential $V_{1}^{\prime}$. If as before contacts 1 and 4 are chosen to be the current contacts then no current flows into the other voltage probe contacts. Lets say a current $\frac{e^2}{h} (T_{2} V_{1}^{\prime}+T_{2} V_{2}^{\prime})$ enters contact 2, the first part$ \frac{e^2}{h} T_{2} V_{1}^{\prime}$ is the spin-up component while the second part  $ \frac{e^2}{h} T_{2} V_{2}^{\prime}$ is the spin-down component moving in exactly the opposite direction.
Similarly, the current $\frac{e^2}{h} 2 T_{2} V_{2}$ leaves contact 2, and since contact 2 is a voltage probe net current has to be zero, implying $V_{2}=(V_{1}^{\prime}+V_{2}^{\prime})/2$. The same thing happens at contact 3 and along the lower edge.

Now we write the current voltage relations in continuous fashion, eschewing our earlier method of writing it in matrix form to avoid clutter as there are not only the 6 potentials $V_{1}-V_{6}$, we also have the equilibrated potentials $V_{1}^{\prime}-V_{6}^{\prime}$.

\begin{eqnarray}
I_{1}&=& T_{1}(2V_{1}-V_{1}^{\prime}-V_{6}^{\prime}),\nonumber\\
I_{2}&=& T_{2}(2V_{2}-V_{1}^{\prime}-V_{2}^{\prime}),\nonumber\\
I_{3}&=& T_{3}(2V_{3}-V_{2}^{\prime}-V_{3}^{\prime}),\nonumber\\
I_{4}&=& T_{4}(2V_{4}-V_{3}^{\prime}-V_{4}^{\prime}),\nonumber\\
I_{5}&=& T_{5}(2V_{5}-V_{4}^{\prime}-V_{5}^{\prime}),\nonumber\\
I_{6}&=& T_{6}(2V_{6}-V_{5}^{\prime}-V_{6}^{\prime}).
\end{eqnarray}
By putting the condition of net current into voltage probe contacts $2,3,4,5$ to be zero we get the following relations between the contact potentials: $V_{2}=(V_{1}^{\prime}+V_{2}^{\prime})/2, V_{3}=(V_{2}^{\prime}+V_{3}^{\prime})/2, V_{5}=(V_{4}^{\prime}+V_{5}^{\prime})/2, \mbox{and} V_{6}=-V_{5}^{\prime} $.

Further, due to the equilibration the net spin-up current out of contact 2 is the sum  $\frac{e^2}{h} (T_{2} V_{2}^{}+R_{2} V_{1}^{\prime})$ and the net spin-down current out of contact 3 is the sum  $\frac{e^2}{h} (T_{3} V_{3}^{}+R_{3} V_{3}^{\prime})$ and this should be equal to   $\frac{e^2}{h} 2 M V_{2}^{\prime}$ which is the net current out of $V_{2}^{\prime}$- the equilibrated potential due to inelastic scattering between contacts 2 and 3. Similarly we can write the net spin polarized currents into and out of the equilibrated potentials. Since there are 6 equilibrated potentials we will have six such equations. 
  The origin of the first equation has already been explained above herein below we  list all of them:
\begin{eqnarray}
\frac{e^2}{h} (T_{2} V_{2}^{}+R_{2} V_{1}^{\prime}) + \frac{e^2}{h} (T_{3} V_{3}^{}+R_{3} V_{3}^{\prime}) &=&\frac{e^2}{h} 2 M  V_{2}^{\prime}\nonumber\\
\frac{e^2}{h} (T_{1} V_{1}^{}+R_{1} V_{6}^{\prime}) + \frac{e^2}{h} (T_{2} V_{2}^{}+R_{2} V_{2}^{\prime}) &=&\frac{e^2}{h} 2 M  V_{1}^{\prime}\nonumber\\
\frac{e^2}{h} (T_{3} V_{3}^{}+R_{3} V_{2}^{\prime} )+ \frac{e^2}{h} (T_{4} V_{4}^{}+R_{4} V_{4}^{\prime})&=&\frac{e^2}{h} 2 M  V_{3}^{\prime}\nonumber\\
\frac{e^2}{h} (T_{4} V_{4}^{}+R_{4} V_{3}^{\prime} )+ \frac{e^2}{h} (T_{5} V_{5}^{}+R_{5} V_{5}^{\prime})&=&\frac{e^2}{h} 2 M  V_{4}^{\prime}\nonumber\\
\frac{e^2}{h} (T_{5} V_{5}^{}+R_{5} V_{4}^{\prime} )+ \frac{e^2}{h} (T_{6} V_{6}^{}+R_{6} V_{6}^{\prime})&=&\frac{e^2}{h} 2 M  V_{5}^{\prime}\nonumber\\
\frac{e^2}{h} (T_{1} V_{1}^{}+R_{1} V_{1}^{\prime} )+ \frac{e^2}{h} (T_{6} V_{6}^{}+R_{6} V_{5}^{\prime})&=&\frac{e^2}{h} 2 M  V_{6}^{\prime}.
\end{eqnarray}

Solving the above six equations, gives the equilibrated potentials $V_{i}^{\prime}, i=1,..6$ in terms of the contact potentials  $V_{i}^{}, i=1,..6$. Substituting the obtained $V_{i}^{\prime}, i=1,..6$ in Eq. ~12, we can derive the necessary resistances.  We specially consider the case wherein $D_{c}=D_{1}=D_{4}\neq D_{u}=D_{2}=D_{3} \neq D_{l}=D_{5}=D_{6}$ and we have-
\begin{eqnarray}
R_{H}&=&R_{14,26}=\frac{h}{2 e^{2} M}\frac{(D_{l}-D_{u})}{3+2D_{l}+D_{u}(2+D_{l})+D_{c}(2+D_{l}+D_{u})}, \nonumber\\
R_{L}&=&R_{14,23}=\frac{h}{ e^{2} M}\frac{(3+2D_{c}+D_{l})}{3+2D_{l}+D_{u}(2+D_{l})+D_{c}(2+D_{l}+D_{u})}\nonumber\\
R_{2T}&=&R_{14,14}=\frac{h}{ e^{2}M}\frac{(4D_{c}^{2}-(3+D_{l})(3+D_{u}))}{(-1+D_{c})(3+2D_{l}+D_{u}(2+D_{l})+D_{c}(2+D_{l}+D_{u}))}.
\end{eqnarray}
Here $D_c$ denotes disorder in current contacts while $D_{u} (D_{l})$ represent disorder in contacts at upper(lower) edge.  When disorder in contacts at upper and lower edge are unequal one sees finite charge Hall conductance and thus pure QSH effect vanishes. This is unlike what happens in this case for QH edge modes not only QH conductance is resilient to disorder and inelastic scattering it retains its quantization and the longitudinal resistance, a measure of voltage drop  across the sample, remains zero. So unlike in case of QH edge modes where inelastic scattering has no effect on Hall quantization in presence of all disordered contacts regardless of whether their strengths are equal or not, in case of QSH effect- inelastic scattering destroys the pure QSH effect in presence of unequal disorder.  Only when contact disorder at various probes are same does pure QSH effect reappear.

In Fig. 3 we plot the Longitudinal and Hall resistances  for QH and QSH cases. One can see from the insets how the spin Hall case is dependent on disorder while QH case remains untroubled by disorder.  

\subsection{QSH edge modes- all disordered probes with inelastic scattering ( without spin flip)}

\begin{figure*}[h]
  \centering
 \subfigure[The case of inelastic scattering without spin-flip. Spin-up edge modes are equilibrated to the primed potentials while spin-down edge modes are equilibrated to the double-primed potentials]{ \includegraphics[width=0.65\textwidth]{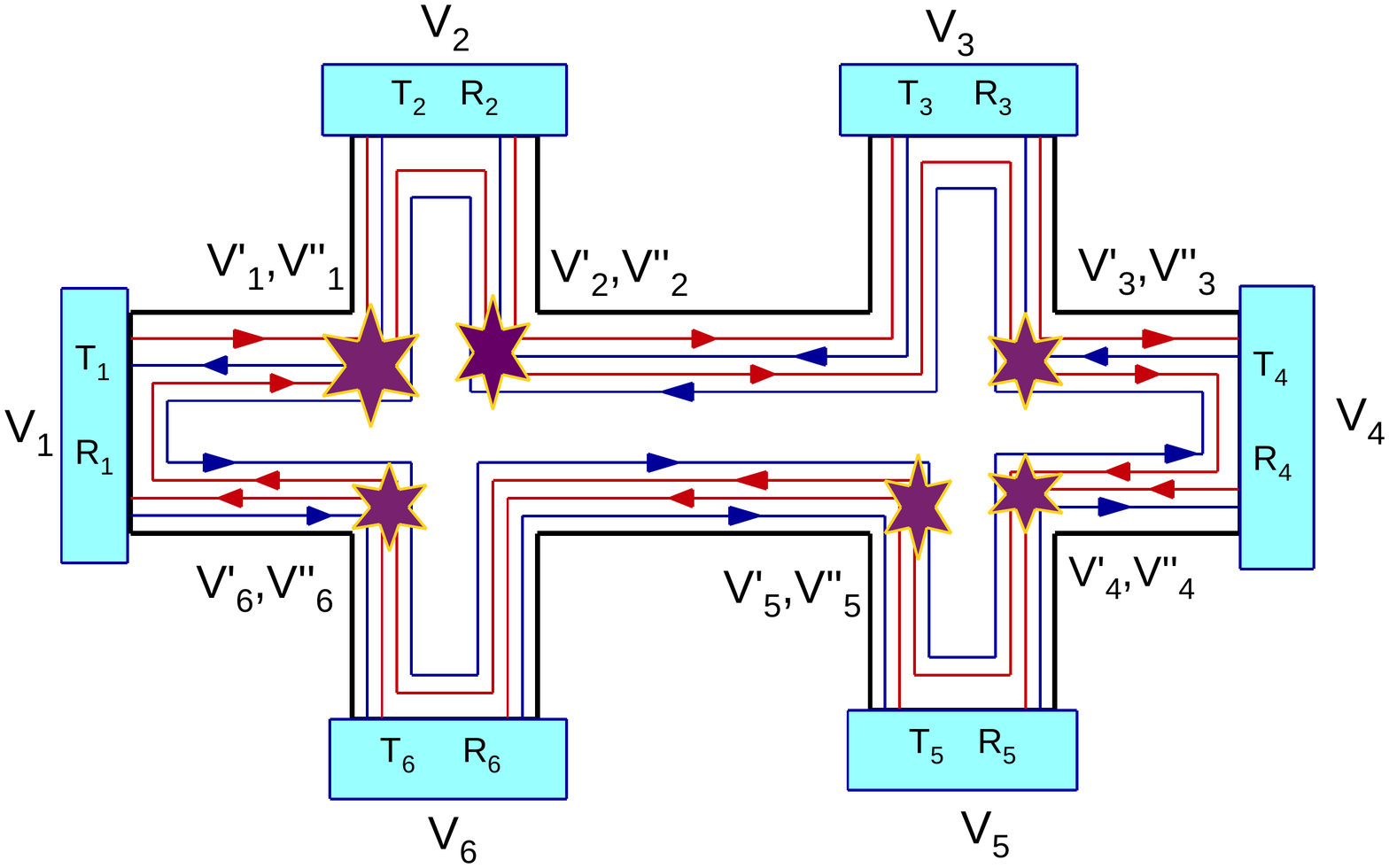}}
  \centering\subfigure  [$R_{H}, R_{L}$ and $R_{2T}$ versus disorder with inelastic and spin-flip scattering included]
{    \includegraphics[width=.49\textwidth]{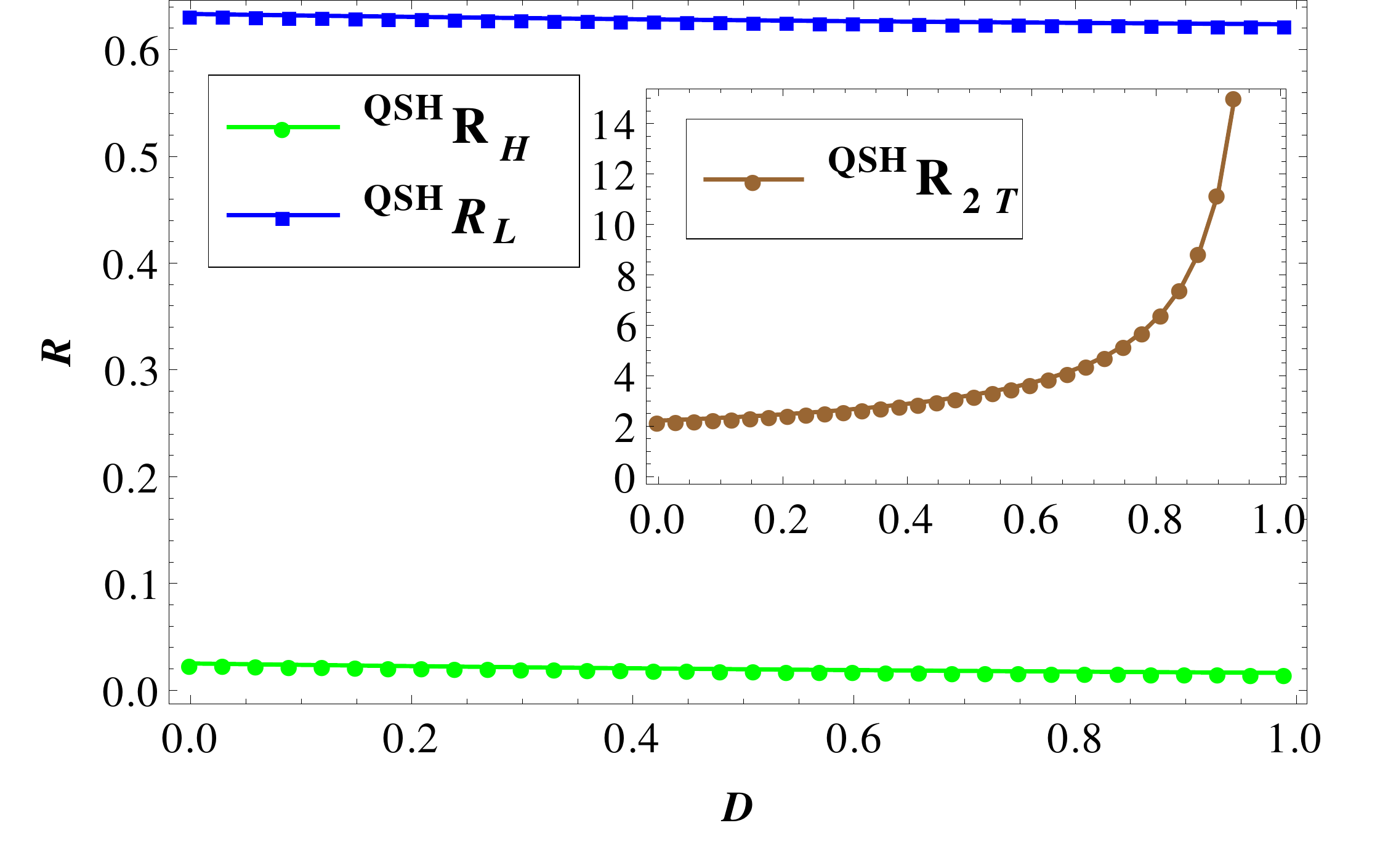}}
 \centering\subfigure [$R_{H}, R_{L}$ and $R_{2T}$ versus disorder with inelastic but without any spin-flip scattering. Curiously resistances are more susceptible to disorder in this case.]{ \includegraphics[width=.49\textwidth]{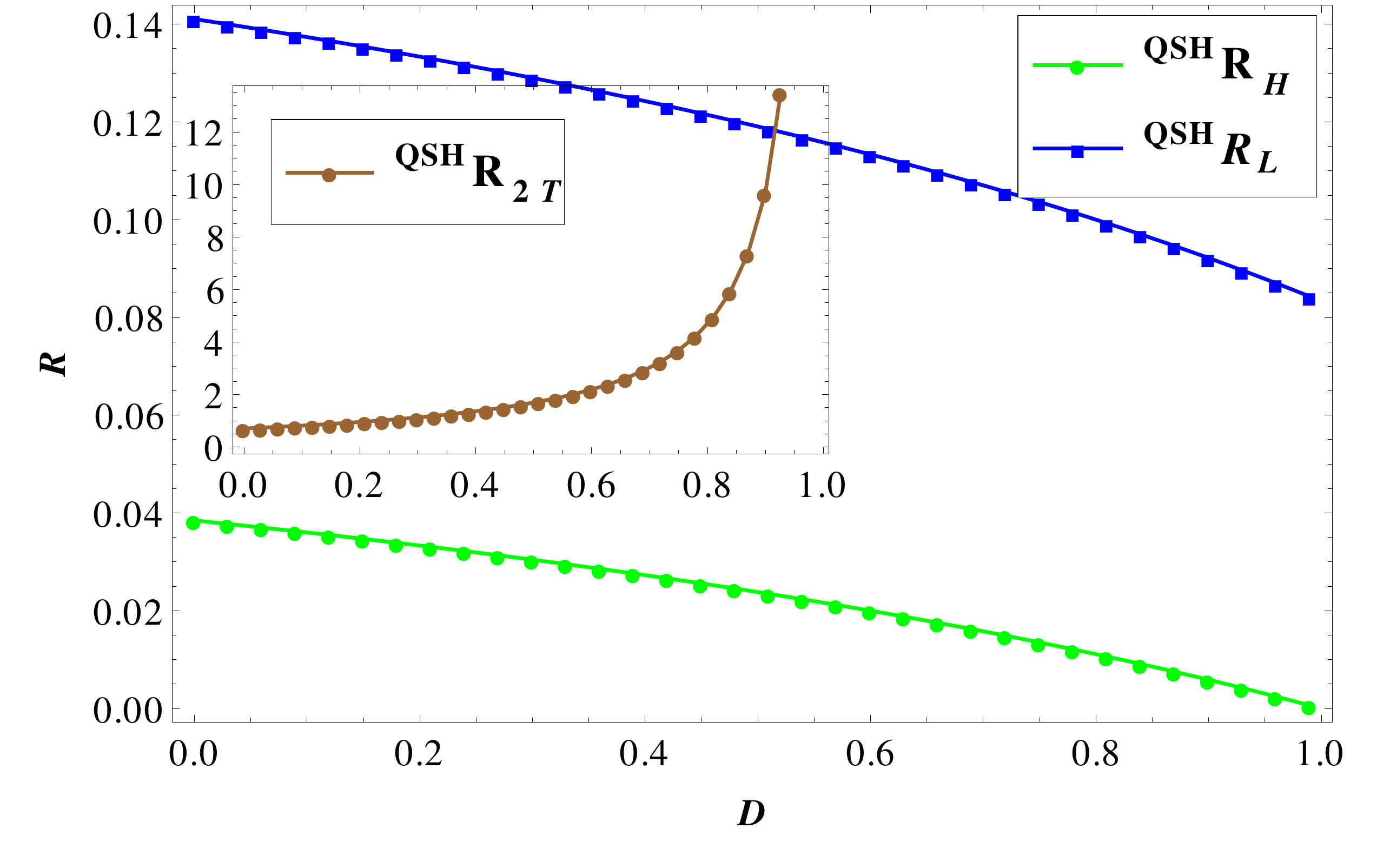}}\caption{The without spin-flip scattering case}
\end{figure*}

The case of QSH edge modes in presence of completely disordered contacts and with inelastic scattering included but now without any spin-flip scattering can be understood by extending the approach of the previous subsection. We can look at the Figure 4(a) where we consider that the length between disordered contacts is larger than inelastic scattering length further we now only have equilibration between same spin edge modes and as before no equilibration occurs across the sample edges. In the event of inelastic scattering  happening, the edge states originating from different contacts with different energies are equilibrated to a common potential as in QH case. In Fig. 4(a), one can see that electrons with spin-up coming from contacts 1 and 6 are equilibrated to potential $V_{1}^{\prime}$ while spin-down electrons coming from contact 2 and 3 are equilibrated to potential $V_{1}^{\prime \prime}$. Similarly, the other potentials $V_{i}^{\prime}, i=2..6$ are decided for equilibration of spin-up edge modes while the potentials $V_{i}^{\prime\prime}, i=2..6$ are decided for equilibration of spin-down edge modes. If as before contacts 1 and 4 are chosen to be the current contacts then no current flows into the other voltage probe contacts. Lets say a spin-up current $\frac{e^2}{h} (T_{2} V_{1}^{\prime}$ and a spin down current $\frac{e^2}{h} (T_{2} V_{2}^{\prime\prime})$ enters contact 2, while the current $\frac{e^2}{h} 2 T_{2} V_{2}$ leaves contact 2, and since contact 2 is a voltage probe net current has to be zero, implying $V_{2}=(V_{1}^{\prime}+V_{2}^{\prime\prime})/2$. The same thing happens at the other contacts.

Now we write the current voltage relations in continuous fashion,as done in previous subsections.  There are not only the 6 potentials $V_{1}-V_{6}$, we also have the equilibrated spin-up potentials $V_{1}^{\prime}-V_{6}^{\prime}$ and the spin down potentials $V_{1}^{\prime\prime}-V_{6}^{\prime\prime}$.

\begin{eqnarray}
I_{1}&=&\frac{e^2}{h}  T_{1}(2V_{1}-V_{1}^{\prime\prime}-V_{6}^{\prime}),\nonumber\\
I_{2}&=&\frac{e^2}{h}  T_{2}(2V_{2}-V_{2}^{\prime\prime}-V_{1}^{\prime}),\nonumber\\
I_{3}&=&\frac{e^2}{h}  T_{3}(2V_{3}-V_{3}^{\prime\prime}-V_{2}^{\prime}),\nonumber\\
I_{4}&=&\frac{e^2}{h}  T_{4}(2V_{4}-V_{4}^{\prime\prime}-V_{3}^{\prime}),\nonumber\\
I_{5}&=&\frac{e^2}{h}  T_{5}(2V_{5}-V_{5}^{\prime\prime}-V_{4}^{\prime}),\nonumber\\
I_{6}&=&\frac{e^2}{h}  T_{6}(2V_{6}-V_{6}^{\prime\prime}-V_{5}^{\prime}).
\end{eqnarray}
By putting the condition of net current into voltage probe contacts $2,3,4,5$ to be zero we get the following relations between the contact potentials: $V_{2}=(V_{1}^{\prime}+V_{2}^{\prime\prime})/2, V_{3}=(V_{2}^{\prime}+V_{3}^{\prime\prime})/2, V_{5}=(V_{4}^{\prime}+V_{5}^{\prime\prime})/2, \mbox{and}   V_{6}=(V_{5}^{\prime}+V_{6}^{\prime\prime})/2$.

Further, due to the equilibration the net spin-up current out of contact 1 is the sum  $\frac{e^2}{h} (T_{1} V_{1}^{}+R_{1} V_{6}^{\prime})$ and this should be equal to $\frac{e^2}{h} M V_{1}^{\prime}$ which is the net current out of the spin-up equilibrated potential $V_{1}^{\prime}$. Similarly, the net spin up currents out of contacts 2-6 are equilbrated to the potentials $V_{i}^{\prime}, i=2..6$ see Eq.16. 

The same procedure we adopt for the down spin currents and these are written below in Eq. 16.
  The origin of the first equation has already been explained above herein below we  list all of them:
\begin{eqnarray}
{\frac{e^2}{h} (T_{1} V_{1}^{}+R_{1} V_{6}^{\prime})=\frac{e^2}{h} M  V_{1}^{\prime}}\nonumber  \qquad \qquad   &&   \qquad\qquad    {\frac{e^2}{h} (T_{1} V_{1}^{}+R_{1} V_{1}^{\prime\prime})=\frac{e^2}{h}MV_{6}^{\prime\prime}}\nonumber \\
{\frac{e^2}{h} (T_{2} V_{2}^{}+R_{2} V_{1}^{\prime})=\frac{e^2}{h}M  V_{2}^{\prime}}  \nonumber \qquad \qquad   &&   \qquad\qquad    {\frac{e^2}{h} (T_{2} V_{2}^{}+R_{2} V_{2}^{\prime\prime})=\frac{e^2}{h}MV_{1}^{\prime\prime}}\nonumber\\
{\frac{e^2}{h} (T_{3} V_{3}^{}+R_{3} V_{2}^{\prime})=\frac{e^2}{h}M  V_{3}^{\prime}} \nonumber \qquad \qquad   &&   \qquad\qquad    {\frac{e^2}{h} (T_{3} V_{3}^{}+R_{3} V_{3}^{\prime\prime})=\frac{e^2}{h}MV_{2}^{\prime\prime}}\nonumber\\
{\frac{e^2}{h} (T_{4} V_{4}^{}+R_{4} V_{3}^{\prime})=\frac{e^2}{h}M  V_{4}^{\prime}}  \nonumber\qquad \qquad   &&   \qquad\qquad    {\frac{e^2}{h} (T_{4} V_{4}^{}+R_{4} V_{4}^{\prime\prime})=\frac{e^2}{h}MV_{3}^{\prime\prime}}\nonumber \\
{\frac{e^2}{h} (T_{5} V_{5}^{}+R_{5} V_{4}^{\prime} )=\frac{e^2}{h}M  V_{5}^{\prime}}  \nonumber \qquad \qquad   &&   \qquad\qquad    {\frac{e^2}{h} (T_{5} V_{5}^{}+R_{5} V_{5}^{\prime\prime})=\frac{e^2}{h}MV_{4}^{\prime\prime}}\nonumber\\
 {\frac{e^2}{h} (T_{6} V_{6}^{}+R_{6} V_{5}^{\prime})=\frac{e^2}{h}M  V_{6}^{\prime}} \nonumber \qquad \qquad   &&   \qquad\qquad    {\frac{e^2}{h} (T_{6} V_{6}^{}+R_{6} V_{6}^{\prime\prime})=\frac{e^2}{h}MV_{5}^{\prime\prime}}\\
\end{eqnarray}

Solving the above twelve equations, gives the equilibrated potentials $V_{i}^{\prime} \mbox { and } V_{i}^{\prime\prime} i=1,..6$ in terms of the contact potentials  $V_{i}^{}, i=1,..6$. Substituting the obtained $V_{i}^{\prime} \mbox{and } V{i}^{\prime\prime}, i=1,..6$ in Eq. ~15, we can derive the necessary resistances.   We have-
\begin{eqnarray}
R_{H}&=&R_{14,26}=\frac{h}{2 e^{2} M}\frac{(-1+D_{c})(D_{l}-D_{u})}{-3+D_{l}(-1+D_{u})-D_{u}+D_{c}(-1+D_{l}+D_{u}+3D_{l}D_{u})}, \nonumber\\
R_{L}&=&R_{14,23}=\frac{h}{2 e^{2} M}\frac{(3+D_{c}(1-3D_{l})-D_{l})(-1+D_{u})}{-3+D_{l}(-1+D_{u})-D_{u}+D_{c}(-1+D_{l}+D_{u}+3D_{l}D_{u})}\nonumber\\
R_{2T}&=&R_{14,14}=\frac{h}{2 e^{2}M}\frac{(3-D_{l})(3-D_{u})+D_{c}^{2}(-1+D_{l}(3-9D_{u})+3D_{u})}{(-1+D_{c})(-3+D_{l}(-1+D_{u})-D_{u}+D_{c}(-1+D_{l}+D_{u}+3D_{l}D_{u})}.
\end{eqnarray}
Here $D_c$ denotes disorder in current contacts while $D_{u} (D_{l})$ represent disorder in contacts at upper(lower) edge.  When disorder in contacts at upper and lower edge are unequal one sees finite charge Hall conductance and thus pure QSH effect vanishes. This is unlike what happens in this case for QH edge modes not only QH conductance is resilient to disorder and inelastic scattering it retains its quantization and the longitudinal resistance a measure of voltage drop  across the sample remains zero. So unlike in case of QH edge modes where inelastic scattering has no effect on the Hall quantization in presence of all disordered contacts regardless of whether their strengths are equal or not, in case of QSH edge modes inelastic scattering destroys the pure QSH effect in presence of unequal disorder.  Only when contact disorder at various probes are same does pure QSH effect reappear.

In Fig. 4 (b) we plot the Longitudinal, Hall and 2Terminal resistances for QSH case with spin-flip and in 4(c) we plot the same without spin-flip. One can see that for the without spin flip scattering case, disorder has a more dramatic effect on the resistances than for that with spin-flip. Somehow inelastic and spin-flip scattering combine to reduce the effect of disorder on the resistances.

\section{Generalization to N terminals}
\begin{figure*}
 \centering \subfigure[QH- ideal case]{ \includegraphics[width=0.45\textwidth]{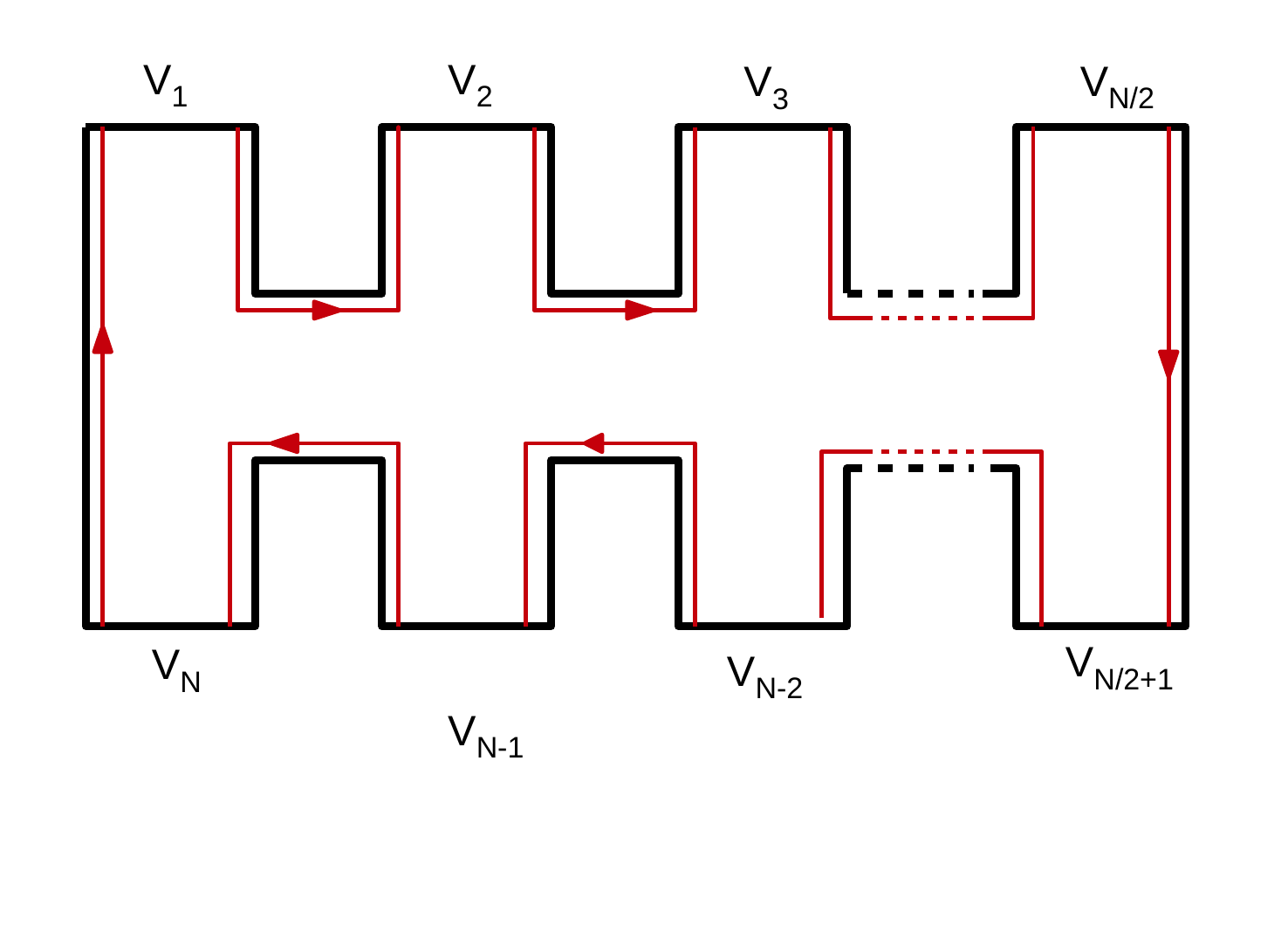}}
 \centering\subfigure[QH-All disordered contacts with inelastic scattering]{   \includegraphics[width=.5\textwidth]{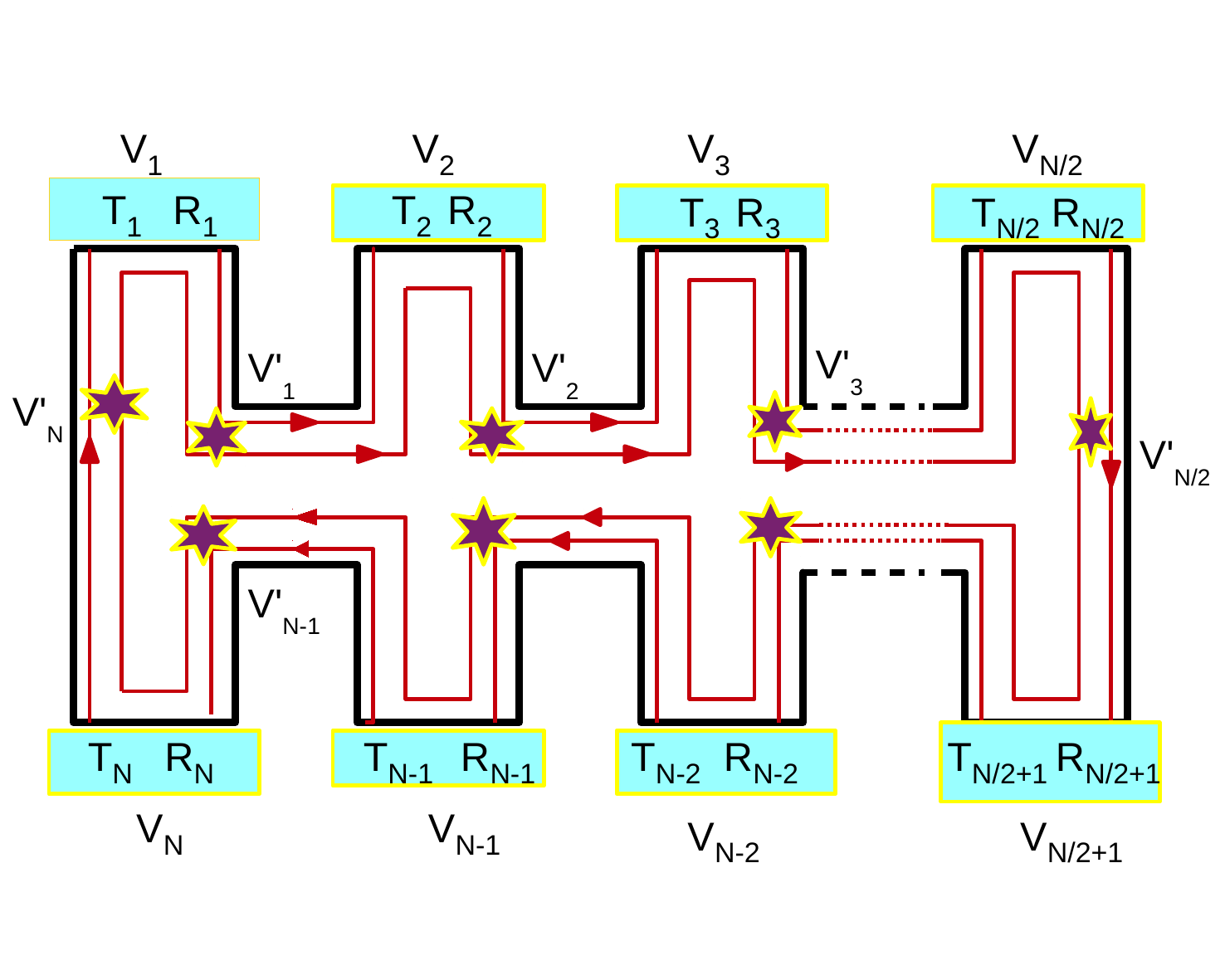}}\caption{N terminal QH bar showing QH edge modes}
\end{figure*}
\subsection{N Terminal QH bar}
\subsubsection{Ideal case}
The ideal case is represented in Fig.~5(a). The current voltage relations can be derived from the current voltage equation:

\begin{equation}
 \left( \begin{array}{c}
    I_{1} \\
    I_{2}  \\ .. \\ I_{k-1} \\ I_{k}\\I_{k+1}\\..\\I_{N-1} \\I_{N}\end{array} \right)=-\frac{e^{2} M}{h} \left( \begin{array}{ccccccccc}
    -1  & 0 & .. & 0 &0 & 0& ..&0 &1 \\
    1  &-1 & .. & 0 &0 & 0&.. &0 &0 \\ .. & .. & .. & .. &.. & ..& ..& ..& ..\\ 0  & 0 & .. & -1 &0 & 0&.. &0 &0 \\ 0 & 0 & .. & 1 &-1 & 0&.. & 0&0 \\ 0  & 0 & .. & 0 &1 & -1&.. &0 &0 \\ ..  & .. & .. & .. &.. & ..& ..& ..&.. \\ 0 & 0 & .. & 0 &0 & 0&.. & -1& 0\\ 0  & 0 & .. & 0 &0 & 0& ..&1 &-1 \end{array} \right)  \left( \begin{array}{c}
    V_{1} \\
    V_{2}  \\ .. \\ V_{k-1} \\ V_{k}\\V_{k+1}\\..\\V_{N-1} \\V_{N}\end{array} \right)
\end{equation}

Substituting $I_{2}, I_{3}, I_{k-1}$ and $ I_{k+1},..I_{N}=0$ and choosing reference potential $V_{N}=0$, we derive $V_{1}=V_{2}=...=V_{k-1}$ and $V_{k}=...=V_{N-1}=V_{N}=0$. So, the Hall resistance $R_{H}=R_{1k,ij}=\frac{h}{e^{2}}\frac{1}{M}, \mbox
{with} 1 \le i < k \le j \le N$, then  longitudinal resistance $R_{L}=R_{1k,ij}=0, \mbox
{with } 1 \le i, j < k \mbox {  or  } k \le i, j \le N$, and finally two terminal resistance  $R_{2T}=R_{1k,1k}=\frac{h}{e^{2}}\frac{1}{M} \mbox{with  } 1 \le i,j < N$.

The case of a single disordered probe as was done for 6 terminal case can be easily calculated and the resistances-Hall, Longitudinal and two terminal are identical to the ideal case with disorder having no impact.  Importantly the quantization of Hall resistance is independent of any asymmetry in number of contacts at upper and lower edge while as we will see below in case of QSH edge modes this is not the case.
\subsubsection{ All disordered probes with inelastic scattering}
The completely disordered case with inelastic scattering is represented in Fig. 5(b). Here the voltage probe contacts are  disordered, i.e., $D_{i}\ne D_{j}, i, j=1...N {\mbox and} i \ne j$. It does not matter whether the contacts are equally disordered or not.
The disorder strengths $D$'s can be written in terms of the T's- the no. of  transmitted edge modes and M- the total no. of edge modes. So, $T_{i}=(1-D_{i})M$ and $R_{i}=D_{i}M$
We can look at the Figure 5(b) where we consider that the length between disordered contacts is larger than inelastic scattering length. On the occasion of an inelastic scattering event happening the edge states originating from different contacts with different energies are equilibrated to a common potential. In Fig. 5(b), one can see that electrons coming from contact 1 and N are equilibrated to potential $V_{1}^{\prime}$. If as before contacts 1 and k are chosen to be the current contacts then no current flows into the other voltage probe contacts. Lets say a current $\frac{e^2}{h} T_{2} V_{1}^{\prime}$ enters contact 2 while current $\frac{e^2}{h} T_{2} V_{2}$ leaves contact 2, and since contact 2 is a voltage probe net current has to be zero, implying $V_{2}=V_{1}^{\prime}$. The same thing happens at contact 3 and along the lower edge where states are equilibrated to $V_{k}^{\prime}$.

Now we write the current voltage relations in continuous fashion, eschewing our earlier method of writing it in matrix form to avoid clutter as there are not only the N potentials $V_{1}-V_{N}$, we also have the equilibrated potentials $V_{1}^{\prime}-V_{N}^{\prime}$.
\begin{eqnarray}
I_{1}&=& T_{1}(V_{1}-V_{N}^{\prime}),\nonumber\\
I_{2}&=& T_{2}(V_{2}-V_{1}^{\prime}),\nonumber\\
I_{3}&=& T_{3}(V_{3}-V_{2}^{\prime}),\nonumber\\
..&=&..(..-..), \nonumber\\
I_{k-1}&=& T_{k-1}(V_{k-1}-V_{k-2}^{\prime}),\nonumber\\
I_{k}&=& T_{k}(V_{k}-V_{k-1}^{\prime}),\nonumber\\
I_{k+1}&=& T_{k+1}(V_{k+1}-V_{k}^{\prime}),\nonumber\\
..&=&..(..-..), \nonumber\\
I_{N}&=& T_{N}(V_{N}-V_{N-1}^{\prime}).
\end{eqnarray}
By putting the condition of net current into voltage probe contacts $2,3,...k-1, k+1...,N$ to be zero we get the following relations between the contact potentials: $V_{2}=V_{1}^{\prime}, V_{3}=V_{2}^{\prime},.... V_{k-1}=V_{k-2}^{\prime}, \mbox{and} V_{k+1}=V_{k}^{\prime},...,V_{N}=V_{N-1}^{\prime}$.
Further, due to the equilibration the net current just out of contact 2 is the sum  $\frac{e^2}{h} (T_{2} V_{2}+R_{2} V_{1}^{\prime})$ and this should be equal to   $\frac{e^2}{h} M V_{2}^{\prime}$ which is the equilibrated potential due to inelastic scattering between contacts 2 and  3.

Thus, $\frac{e^2}{h} (T_{2} V_{2}+R_{2} V_{2})=\frac{e^2}{h} M V_{2}^{\prime}$, or $V_{2}=V_{2}^{\prime}$, as $T_{2}+R_{2}=M$ the total no. of edge modes in the system. Thus all the upper edges are equilibrated to same potential $V_{1}^{\prime}=V_{2}=V_{2}^{\prime}=V_{3}=V_{3}^{\prime}=V_{1}=...=V_{k-1}=V_{k-1}^{\prime}$.

  Similarly for the equilibrated potentials at the lower edge we get $V_{k}^{\prime}=V_{k+1}=V_{k+1}^{\prime}=...=V_{N}=V_{N}^{\prime}=0$, as $V_{N}=0$. So, the Hall resistance $R_{H}=R_{1k,ij}=\frac{h}{e^{2}}\frac{1}{M}  \mbox{  with } 1 < i < k < j \le N$,   longitudinal resistance $R_{L}=R_{1k,ij}=0  \mbox{  with } 1 < i , j < k  \mbox{  or  } k < i , j \le N$, and two terminal resistance  $R_{2T}=R_{1k,1k}=\frac{h}{e^{2}} \frac{M^{2}-R_{1}R_{k}}{MT_{1}T_{k}}.$ 
Thus sample geometry  has no role in this case.
\begin{figure*}
  \centering
 \subfigure[N terminal QSH bar-ideal case]{ \includegraphics[width=0.45\textwidth]{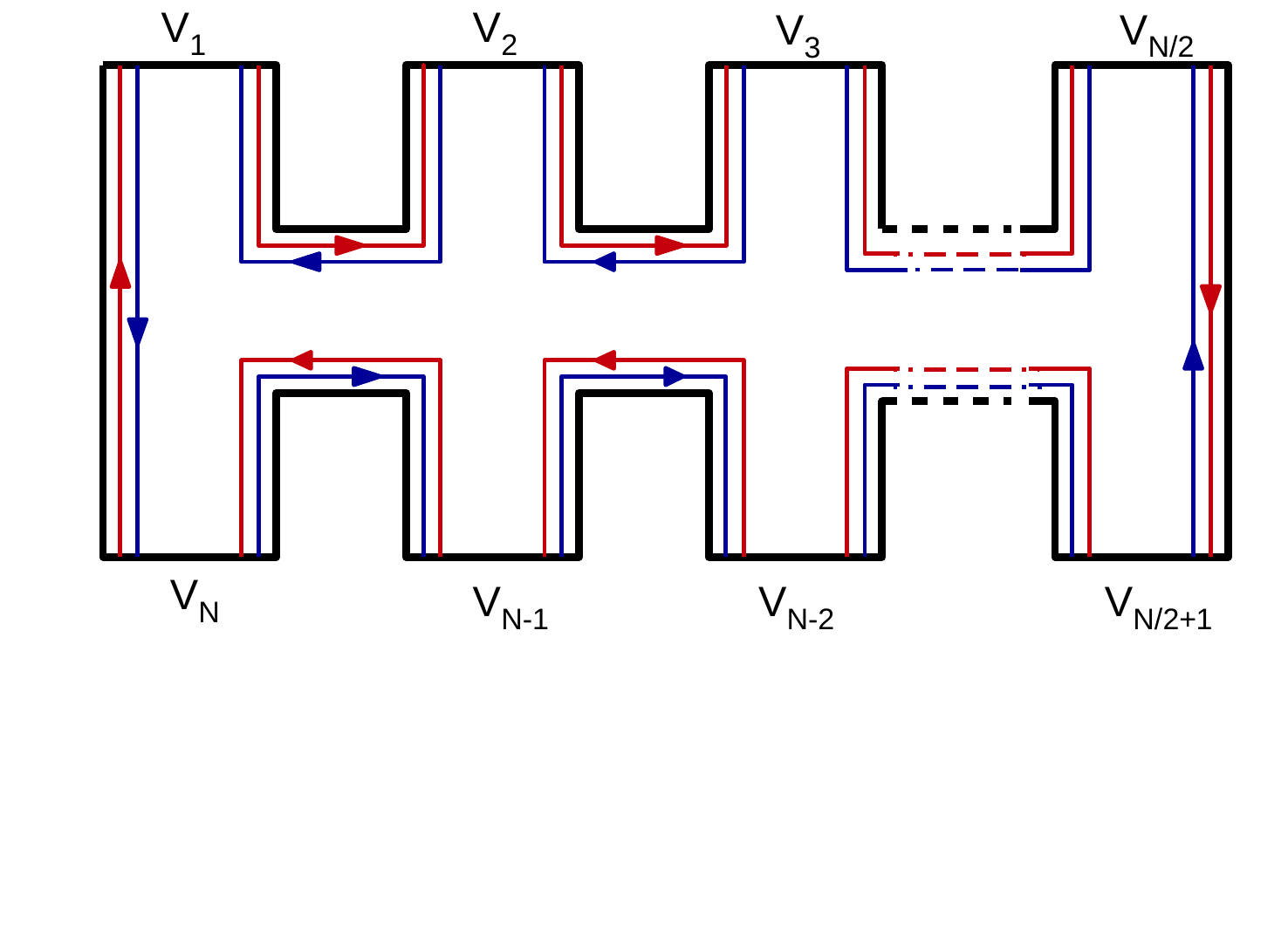}}
   \centering
\subfigure[N terminal QSH bar-All disordered contacts with inelastic scattering]{   \includegraphics[width=.5\textwidth]{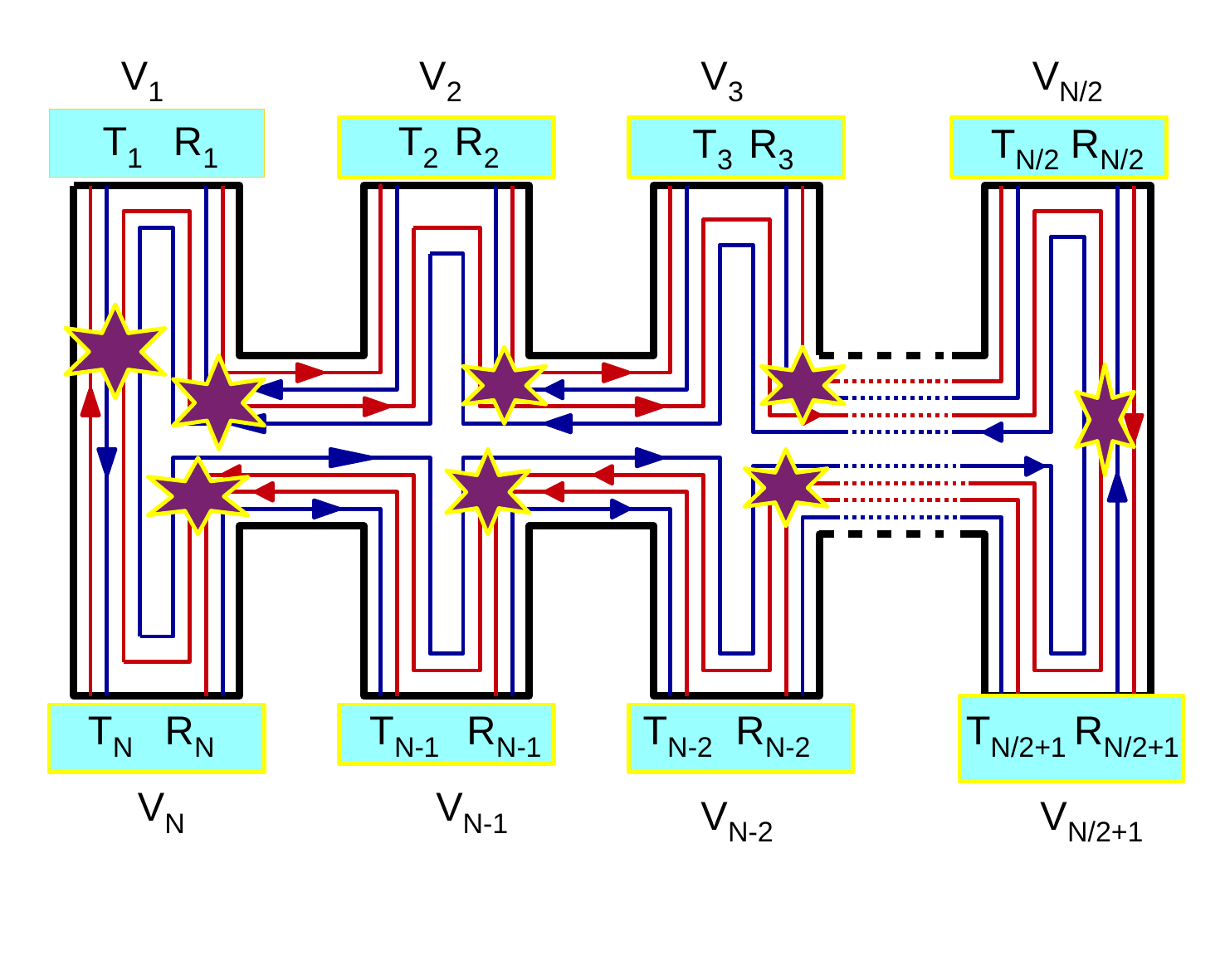}}\caption{N terminal QSH bar showing  edge modes.}
\end{figure*} 
\subsection{N Terminal QSH bar}
\subsubsection{Ideal case}
The ideal case is represented in Fig.6(a). The current voltage relations can be derived from the current voltage equation:
\begin{equation}
 \left( \begin{array}{c}
    I_{1} \\
    I_{2}  \\ .. \\ I_{k-1} \\ I_{k}\\I_{k+1}\\..\\I_{N-1} \\I_{N}\end{array} \right)=-\frac{e^{2} M}{h} \left( \begin{array}{ccccccccc}
    -2  & 1 & .. & 0 &0 & 0& ..&0 &1 \\
    1  &-2 & .. & 0 &0 & 0&.. &0 &0 \\ .. & .. & .. & .. &.. & ..& ..& ..& ..\\ 0  & 0 & .. & -2 &1 & 0&.. &0 &0 \\ 0 & 0 & .. & 1 &-2 & 1&.. & 0&0 \\ 0  & 0 & .. & 0 &1 & -2&.. &0 &0 \\ ..  & .. & .. & .. &.. & ..& ..& ..&.. \\ 0 & 0 & .. & 0 &0 & 0&.. & -2& 1\\ 1  & 0 & .. & 0 &0 & 0& ..&1 &-2 \end{array} \right)  \left( \begin{array}{c}
    V_{1} \\
    V_{2}  \\ .. \\ V_{k-1} \\ V_{k}\\V_{k+1}\\..\\V_{N-1} \\V_{N}\end{array} \right)
\end{equation}

Substituting $I_{2}, I_{3}, I_{k-1}$ and $ I_{k+1},..I_{N}=0$ and choosing reference potential $V_{N}=0$, we derive $V_{i}=(i-1)V_{2}-(i-2)V_{1}$, where $1 \le i \le k$ with $V_{2}=\frac{2k-N-2}{k-1} V_{1}$.  Similarly, $V_{i}=-(N-i)V_{1}$, where $k \le i \le N$.

So, the Hall resistance 
\begin{eqnarray}
R_{H}=R_{1k,ij}&=& \frac{h}{e^{2} M} (\frac{(i-1)(2k-N-2)}{N}, \mbox
{with} 1 \le i  \le k, \end{eqnarray}
   longitudinal resistance 
\begin{eqnarray}
R_{L}=R_{1k,ij}&=&\frac{h}{e^{2}M}(\frac{j-i}{N})(1-k+N), \mbox
{with } 1 \le i,j < k \nonumber\\
&=& \frac{h}{e^{2}M}(\frac{j-i}{N})(1-k), \mbox
{with } k \le i,j < N \nonumber
\end{eqnarray}
and two terminal resistance  $R_{2T}=R_{1k,1k}=\frac{h}{e^{2}M}\frac{k-1}{N}(1-k+N), \mbox{with  } 1 \le i,j \le N$.
Surprisingly, a finite charge Hall current flows even when there is no disorder. It arises only due to asymmetry between number of contacts at upper and lower edge. This number asymmetry has no role as far as QH edge modes are concerned.
\subsubsection{All disordered probes with inelastic scattering and spin-flip scattering}
The completely disordered case with inelastic scattering is represented in Fig.~6(b). Here both the voltage as well as current probe contacts are  disordered, i.e., $D_{i}= D_{j}, i, j=1...N$. To simplify the calculation- all the contacts are considered equally disordered.
The disorder strengths $D$'s can be written in terms of the T's- the no. of  transmitted edge modes and M- the total no. of edge modes. So, $T_{i}=T=(1-D_{i})M=(1-D)M$ and $R_{i}=R=D_{i}M=DM$.

In Fig.~6(b), we consider the length between disordered contacts to be larger than the inelastic scattering length. On the occasion of an inelastic scattering event happening the edge states originating from different contacts with different energies are equilibrated to a common potential. In Fig.~6(b), one can see that electrons coming from contact 1 and N are equilibrated to potential $V_{1}^{\prime}$. If as before contacts 1 and k are chosen to be the current contacts then no current flows into the other voltage probe contacts. 
 Lets say a current $\frac{e^2}{h} (T_{2} V_{1}^{\prime}+T_{2} V_{2}^{\prime})$ enters contact 2, the first part$ \frac{e^2}{h} T_{2} V_{1}^{\prime}$ is the spin-up component while the second part  $ \frac{e^2}{h} T_{2} V_{2}^{\prime}$ is the spin-down component moving in exactly the opposite direction.
Similarly, the current $\frac{e^2}{h} 2 T_{2} V_{2}$ leaves contact 2, and since contact 2 is a voltage probe net current has to be zero, implying $V_{2}=(V_{1}^{\prime}+V_{2}^{\prime})/2$. The same thing happens at contact 3 and along the lower edge.

Now we write the current voltage relations in continuous fashion, and as before eschewing our earlier method of writing it in matrix form to avoid clutter as there are not only the N potentials $V_{1}-V_{N}$, we also have the equilibrated potentials $V_{1}^{\prime}-V_{N}^{\prime}$.
\begin{eqnarray}
I_{1}&=& 2T V_{1} -T (V_{1}^{\prime}+V_{N}^{\prime}),\nonumber\\
I_{2}&=& 2 T V_{2}- T(V_{1}^{\prime}+V_{2}^{\prime}),\nonumber\\
I_{3}&=& 2 T V_{3}-T(V_{3}^{\prime}+V_{2}^{\prime}),\nonumber\\
..&=&..(..-..), \nonumber\\
I_{k-1}&=& 2TV_{k-1}-T(V_{k-1}^{\prime}+V_{k-2}^{\prime}),\nonumber\\
I_{k}&=& 2T V_{k}-T(V_{k-1}^{\prime}+V_{k}^{\prime}),\nonumber\\
I_{k+1}&=& 2TV_{k+1}-T(V_{k}^{\prime}+V_{k+1}^{\prime}),\nonumber\\
..&=&..(..-..), \nonumber\\
I_{N}&=& 2TV_{N}-T(V_{N-1}^{\prime}+V_{N}^{\prime}).
\end{eqnarray}
Further, due to the equilibration the net current just out of contact 2 is the sum  $\frac{e^2}{h} (T V_{2}+R V_{1}^{\prime}+R V_{3}^{\prime}+T V_{3})$ and this should be equal to   $\frac{e^2}{h} 2M V_{2}^{\prime}$ which is the equilibrated potential due to inelastic scattering between contacts 2 and  3.
Similarly for the net current just out of k$^{th}$ contact is the sum $\frac{e^2}{h} (T V_{k}+R V_{k-1}^{\prime}+R V_{k+1}^{\prime}+T V_{k+1})$ and this should be equal to   $\frac{e^2}{h} 2M V_{k}^{\prime}$ which is the equilibrated potential due to inelastic scattering between contacts $k$ and  $k+1$.
By putting the condition of net current into voltage probe contacts $2,3,...k-1, k+1...,N$ to be zero we get the following relations between the contact potentials: $V_{i}^{\prime}=(i-1)V_{2}^{\prime}-(i-2)V_{1}^{\prime}$ with $2 \le i \le (k-1)$ and $V_{i}^{\prime}=-(2N-2i-1)V_{N}^{\prime}$ with $k \le i \le N$.

So, the Hall resistance with $j=N-i+2$ is given as- \[R_{H}=R_{1k,ij}=\frac{h}{e^{2}M} \frac{2(i-1)(2k-N-2)}{(1+D)N}  \mbox{  with } 1 < i < k < j \le N\] and   if we consider  $k=N/2+1$, i.e., a symmetric sample (with equal no of contacts at the lower and upper edge) then $R_{H}=0$. So sample geometry (number asymmetry between contacts at upper and lower edge) has a direct bearing on whether one sees a pure spin Hall effect or it is contaminated by a charge current. Further in symmetric case although there is no charge Hall current- it is seen only when all the probes are equally disordered, i.e., QSH effect is restored. If on the other hand probes are not equally disordered as seen in 6 terminal case $R_{H}\neq 0$ even in presence of inelastic scattering. So unlike in case of QH edge modes where inelastic scattering restores Hall quantization in presence of all disordered contacts regardless of their strengths are equal or not, in case of QSH inelastic scattering also fails to restore the pure QSH effect in presence of unequal disorder.

 Next, longitudinal resistance-
\begin{eqnarray}
R_{L}=R_{1k,ij}&=& \frac{h}{e^{2}M}  \frac{(j-i)2(1-k+N)}{(1+D)N} \mbox{  with } 2 < i , j < k \nonumber\\
&=&  \frac{h}{e^{2}M}  \frac{2(i-j)(k-1)}{(1+D)N},  \mbox{  with  } k+1 \le i , j \le N.
\end{eqnarray}
and finally the two terminal resistance  \[R_{2T}=R_{1k,1k}= -\frac{h}{e^{2}M} \frac{2[(k-1)(k-N-1)-D(1+k^{2}+2N-k(N+2))]}{(1-D)(1+D)N}.\]

The N-terminal results reduce to the 6 terminal case of equal number of probes at the upper and lower edge (symmetric case) in all cases confirming the results obtained as before. Further they shed light on the asymmetric case wherein probes on upper and lower edge are unequal. Asymmetry has no role as far as QH edge modes are concerned but in case of QSH edge modes they have a non-trivial role, even destroying the pure QSH effect, regardless of whether there is disorder or not. 
\section{ Conclusions}
The conclusions of the work reported here shed a searchlight on QSH edge modes which hitherto had enjoyed a rather sunny outlook vis a vis QH edge modes. We in this work shatter the myth that QSH edge modes are better in fact we show them to be worse as compared to QH edge modes as far as their ability to  withstand the deleterious effects of disorder and inelastic scattering are concerned. Further, even sample geometry can have a negative bearing on QSH effect while leaving QH effect intact.
Given that QSH insulators are considered to be the next big breakthrough after graphene, their complete and utter failure to withstand disorder and sample asymmetry not to mention inelastic scattering means that they will limited use in spintronics and applications otherwise. Further, since topological insulator edge modes are considered to be useful in a host of other areas ranging from topological quantum computation (braiding of majorana fermions) to searching for novel spin dependent effects, this paper casts a long shadow of doubt as regards their utility. Caution is the watchword and  any benefits from QSH edge modes will have to counter balanced by the implications of our work.   
\acknowledgments
This work was supported by funds from Dept. of Science and Technology (Nanomission), Govt. of India, Grant No. SR/NM/NS-1101/2011.
  
\end{document}